\documentclass[12pt]{article}
\newcommand{\Comment}[1]{{}}
\usepackage{amsfonts,amsthm,amsmath,amssymb,slashed}
\usepackage[textwidth = 430 pt, textheight = 630 pt]{geometry}

\Comment{\usepackage{color}
\definecolor{MyDarkBlue}{rgb}{0.15,0.15,0.45}
\usepackage[linktocpage=true]{hyperref}
\hypersetup{
colorlinks=true,
citecolor=MyDarkBlue,\bibliography{heisenberg}
linkcolor=MyDarkBlue,
urlcolor=MyDarkBlue,
pdfauthor={Horatiu Nastase and Jacob Sonnenschein},
pdftitle={More on Heisenberg model for high energy nucleon-nucleon scattering},
pdfsubject={hep-th}
}

\usepackage[numbers,sort&compress]{natbib}
\usepackage{hypernat}}
\usepackage{graphicx}

\newcommand\ignore[1]{}
\def\one{{\,\hbox{1\kern-.8mm l}}}

\newcommand{\tr}{\operatorname{tr}}

\def\a{\alpha}\def\b{\beta}

\def\r{\rho}\def\k{\kappa}

\def\d{\partial}

\def \pa {\partial}

\newcommand{\Cset}{{\,\,{{{^{_{\pmb{\mid}}}}\kern-.45em{\mathrm C}}}}}

\newcommand{\be}{\begin{equation}}
\newcommand{\bea}{\begin{eqnarray}}

\newcommand{\ee}{\end{equation}}
\newcommand{\eea}{\end{eqnarray}}

\newcommand{\non}{\nonumber \\}
\newcommand{\CR}{\non\cr}

\begin{document}

\renewcommand{\thefootnote}{\fnsymbol{footnote}}

\makeatletter
\@addtoreset{equation}{section}
\makeatother
\renewcommand{\theequation}{\thesection.\arabic{equation}}

\rightline{}
\rightline{}
   \vspace{1.8truecm}


\vspace{10pt}


\begin{center}
{\LARGE \bf{\sc  More on Heisenberg's  model for high energy nucleon-nucleon scattering}}
\end{center}
 \vspace{1truecm}
\thispagestyle{empty} \centerline{
{\large \bf {\sc Horatiu Nastase${}^{a,}$}}\footnote{E-mail address: \Comment{\href{mailto:nastase@ift.unesp.br}}{\tt
    nastase@ift.unesp.br}}
{\bf{\sc and}}
 {\large \bf {\sc Jacob Sonnenschein${}^{b,}$}}\footnote{E-mail address: \Comment{\href{mailto:cobi@post.tau.ac.il}}{\tt cobi@post.tau.ac.il}}
                                                           }

\vspace{1cm}

\vspace{.8cm}
\centerline{{\it ${}^a$
Instituto de F\'{i}sica Te\'{o}rica, UNESP-Universidade Estadual Paulista}} \centerline{{\it
R. Dr. Bento T. Ferraz 271, Bl. II, Sao Paulo 01140-070, SP, Brazil}}

\centerline{{\it ${}^b$
School of Physics and Astronomy,}}
 \centerline{{\it The Raymond and Beverly Sackler Faculty of Exact Sciences, }} \centerline{{\it Tel Aviv University, Ramat Aviv 69978, Israel}}

\vspace{1.0truecm}

\thispagestyle{empty}

\centerline{\sc Abstract}

\vspace{.4truecm}

\begin{center}
\begin{minipage}[c]{380pt}
{\noindent We revisit Heisenberg's model for nucleon-nucleon scattering which admits a  saturation of the Froissart bound.
We examine its uniqueness, and find that up to certain  natural generalizations,
it is the only action that saturates  the bound.
We find that we can extract also sub-leading behaviour for $\sigma_{\rm tot}(s)$ from it,
though that requires a knowledge of the wavefunction solution that is hard to obtain, and a black-disk model allows the calculation of
$\sigma_{elastic}(s)$ as well.

The wavefunction solution is analyzed perturbatively, and its source is interpreted. Generalizations to several mesons,
addition of vector mesons, and curved space regimes are also found.
We discuss the relations between Heisenberg's model and holographic models that are dual to QCD-like theories.

}
\end{minipage}
\end{center}

\vspace{.5cm}

\setcounter{page}{0}
\setcounter{tocdepth}{2}

\newpage

\tableofcontents
\renewcommand{\thefootnote}{\arabic{footnote}}
\setcounter{footnote}{0}

\linespread{1.1}
\parskip 4pt


\section{Introduction}

In quantum field theory  unitarity constraints the asymptotic dependence of the  total cross section  of any scattering process
to be bounded by the well known    Froissart bound
\cite{Froissart:1961ux,Lukaszuk:1967zz},
\be
\sigma_{\rm tot}(s)\leq C\ln^2\frac{s}{s_0};\;\;\; C\leq \frac{\pi}{m^2}\;,
\ee
where  $s$ is Mandelstam's dynamical variable and $m$ is the mass of the lightest particle that can be exchanged by the scattering projectiles.
In the case of QCD  $m=m_\pi$ is the pion mass,  and the bound is supposed to be saturated in the $s\rightarrow\infty$ limit.
However, being that the saturation of the bound is governed by nonperturbative, IR, physics, attempts to describe the saturation of the bound in QCD
have not been successful. Strikingly, nine years before the  discovery of the bound, and in fact  even before the birth  of QCD,
Heisenberg proposed a simple effective model for the maximal
behaviour of $\sigma_{\rm tot}(s)$, in terms of a DBI action for the  pion field, that gives an almost  saturation of the Froissart bound in the case of QCD
\cite{Heisenberg1952}.


The paper of Heisenberg  \cite{Heisenberg1952} includes two revolutionary ideas: (i) Extraction of the dependence of the total cross section on
Mandelstam $s$ variable from the average energy per pion determined from the classical energy density of the scalar field. (ii) Describing
the dynamics of the scalar field using the DBI action. The first idea is obviously very different from the way one usually determines cross section
in perturbation theory. Instead of computing Feynman diagrams of scattering amplitudes and then from the amplitudes determining the
cross section, Heisenberg's proposition is to derive the cross section in a very simple manner from the following relation
\be\label{FBksigma} 
\langle k_0\rangle = \sqrt{s}e^{-m_\pi b_{max}},\rightarrow\qquad \sigma_{tot}=\pi b_{max}^2= \frac{\pi}{m_\pi^2}\log^2\frac{s}{\langle k_0\rangle^2}\;,
\ee
where $\langle k_0\rangle$ is the energy per pion, $m_\pi$ is its mass, $b_{max}$ is the maximal impact parameter for which there is still an
interaction between the two nucleon projectiles. The assumption of the model is that there in an
``effective action" for the scalar field that mediates the interaction from which one can compute $\langle k_0\rangle= \frac{{\cal E}}{n}$ where
${\cal E}$  is the total energy and $n$ is the number of the pions. Thus, the dependence of $\sigma_{tot}$ on $s$ follows from the dependence of
$\langle k_0\rangle$ on $s$. A physical system for which $\langle k_0\rangle$ does not depend on $s$ saturates Froissart's bound.
 The second original idea is to use a non-standard action to describe the dynamics of the pion field. In his paper Heisenberg found that using an
 action of the scalar that is based on ordinary kinetic term and regardless of  what is its potential cannot saturate the Froissart bound. In fact, it
 will yield a constant cross section. However,  using a  DBI action  yields $\langle k_0\rangle \sim \log\frac{s}{m_\pi^2}$. This mild dependence of
 $\langle k_0\rangle$ on $s$ means that the total cross section of the model is close to that of the bound.

\begin{figure}[h!]
\begin{center}
\vspace{3ex}
\includegraphics[width= 100mm]{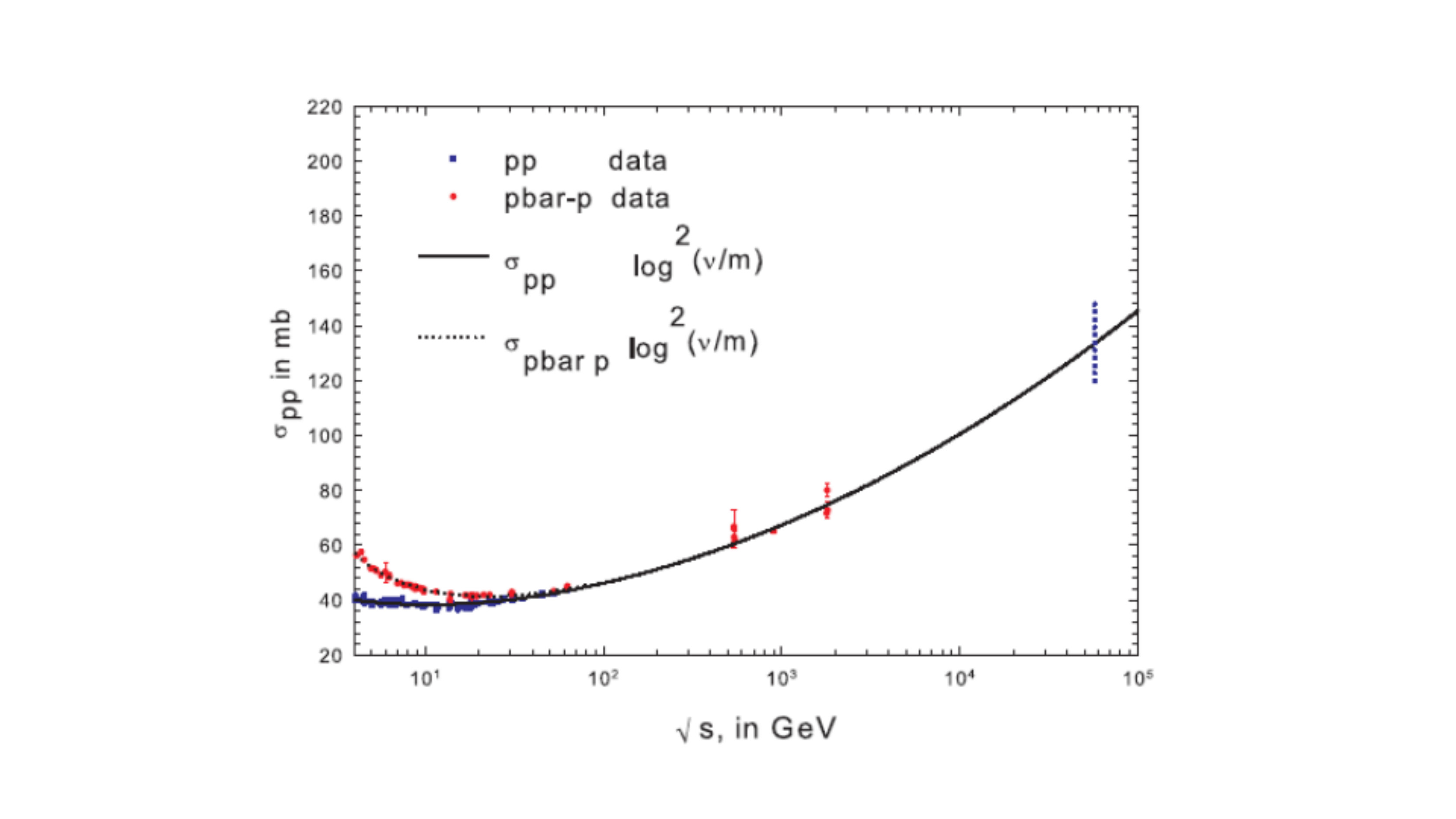}

\end{center}
\caption{The total cross section as a function of $\sqrt{s}$ for $pp$ and $p\bar p$ scattering }
\label{ppscatering}
\end{figure}

Experimental data of the total cross section of proton-proton (and proton- antiproton) collisions is well established on a very wide range of
energies  starting from sub GEV energies and all the way to $\sqrt{s} =7$ in the TOTEM experiment in the LHC and in fact even higher up to
 $\sqrt{s} =57$ TeV from cosmic rays observation. Figure (\ref{ppscatering}) shows the data points together with a fit based on the (\ref{FBksigma})
 but with a mass $m\simeq 1Gev$ and not the pion mass \cite{Nussinov:2008nz}. Thus, regardless of the Froissart bound, one would like to
 have a theoretical model that resembles the behavior of (\ref{FBksigma}) since it seems to fit the experimental data quite well. Needless to
 say that there is no direct derivation from QCD  that can reproduce such a fit.

The goals of this paper are fourfold: (i) Since part of the paper of \cite{Heisenberg1952} is written in a concise form, we elaborate the discussion,
performing several additional  calculations and provide some further evidence for the claims of the paper. In particular we analyze the pion field including the passage from the 1+1 dimensional  solution  to a full four dimensional one.
The ratio of elastic to total cross section is derived using a black disk model.  (ii) We examine the uniqueness of the DBI action as an action that
can (almost) saturate the bound. We prove that only an action with an infinite  tower of higher powers of the derivative term, as the DBI action
admits, can do the job.  (iii) We propose and analyze several generalizations of Heisenberg's model. We add a general potential instead of only
a mass term, and  we  analyze a sigma  model with several scalars.  We examine the ``highly effective action" of \cite{Schwarz:2013wra}for the
case of single scalar in $AdS_5$. For that case the ordinary kinetic term in the square root Lagrangian density undergoes the following
transformation $\pa_\mu\phi\pa^\mu\phi\rightarrow \frac{1}{\phi^4}\pa_\mu\phi\pa^\mu\phi$. Upon considering an $n^{th}$ power rather
than $\phi^4$, we show that only for the range $n\in (0,2)$ we obtain saturation of the Froissart bound. (iv) The last goal has been to
relate Heisenberg's model to the DBI action used in gauge/gravity duality and furthermore to two different holographic  approaches
to the nucleon-nucleon scattering.

One approach to the latter is based on  a simple effective model for QCD scattering at high energies has been developed,
the Polchinski-Strassler model \cite{Polchinski:2001tt} in terms of a metric in a cut-off $AdS_5$ background, dual to glueball fields, and the
fluctuation of an IR brane (IR cut-off), dual to a pion field (the model was extended to the phenomenological "hard-wall" model with the addition
-by hand- of extra fields in the bulk in \cite{Erlich:2005qh}). Based on earlier work in \cite{Giddings:2002cd}, in \cite{Kang:2004jd} it was shown that
the saturation of the Froissart bound can arise through scattering of gravitational shockwaves located on (or close to) the IR brane, with formation of
black hole on the IR brane. Then in \cite{Kang:2005bj} it was shown that one can map exactly the description of the saturation of the bound in the dual,
through gravitational shockwave collision with black hole formation, with the saturation of the bound in the Heisenberg model, through pion field shockwave
collisions. This picture was used further in \cite{Nastase:2005rp,Nastase:2006eb,Freund:2008tv} (see \cite{Nastase:2008hw} for a review) to describe the
sQGP fireball obtained in heavy ion collisions as the object dual to the black hole formed on the IR brane.

In the second approach we relate Heisenberg's DBI action to the DBI that describes the fluctuation of the flavor branes in confining gravitational
background and in particular in  the generalized  Sakai-Sugimoto model \cite{Sakai:2004cn}\cite {Aharony:2006da}. We show that what sources
the scalar field in that model is the flavor instanton density \cite{Hata:2007mb},  that corresponds  to the proton  density.

The paper is organized as follows. In the next section we review the model of \cite{Heisenberg1952} and elaborate on certain issues by
performing additional calculations. In section 3 we discuss the uniqueness of the DBI action
used in \cite{Heisenberg1952} for the saturation of the Froissart bound. We show that  to have a solution of the form $\phi(s)= A\sqrt{s}$,
which is what is needed to saturate the  Froissart bound, one cannot use an action with a finite series of higher derivatives and only the
infinite series that follows from the DBI action does the job (though we were not able to show that some other action with an infinite number of higher
derivative terms couldn't do the job also).
In section 4 we consider various generalization of the action used in \cite{Heisenberg1952}. First we add a potential term in addition to the mass term.
We then analyze a DBI sigma model for several scalar fields. For the simple case of replacing one kinetic term in the square root with  a sum of kinetic
terms for several  scalars the behaviour is similar to the original model. Next we discuss the case of a DBI action associated with
the $AdS_5\times S^5$ case.
We show that for the case of a single scalar field performing the determinant in the DBI action yields a close cousin of action used
in \cite{Heisenberg1952}. Another generalization discussed is  the DBI for  vector mesons. Assuming here again dependence only on the
coordinate $s$ defined in (\ref{defstx}) we show that the behaviour of the vector mesons is similar to that of the scalar meson. In section 5
we discuss the ``wavefunction" of the pion. Firstly we elevate the solution $\phi(s)$ to $\phi(s,r)$.
We argue that for Heisenberg's solution we have a shock-like behaviour, where $T_{++}$ blows up at $x^+=0$, even though we don't have
a $\delta(x^+)$ behaviour. Next we consider a perturbative expansion around $r=0$ and then an pertubative solution around the
asymptotics $r\rightarrow \infty$. Section 6 is devoted to analyzing the sources of the pion field. We first consider sources for  the
nonlinear Born Infeld  theory of electrodynamics. We then in subsection 6.2 discuss in a similar manner the source of a scalar DBI theory.
Section 7 deals with the  original question behind this paper, namely, the cross section of the nucleon-nucleon scattering process.
We discuss  corrections away from the Froissart bound. We then describe the model of the black disk and the corresponding ratio
between the elastic and total cross sections. Section 8 is devoted to an examination of the relation between Heisenberg' model and the
holographic description of nucleon-nucleon scattering. We show that  the nucleon-nucleon scattering process takes the form of  the
scattering   instantons of the flavored gauge fields that reside on the flavor branes. We end this paper in a section of summary and open questions.

\section{  Heisenberg's  model- A review and elaborations }

In this section, we first review the work of Heisenberg in modern language,  and we perform some additional computations that clarify some
aspects of the model.

With a  remarkable insight, Heisenberg considered a nonlinear higher derivative action for the pion, the DBI action with a mass term inside the square root,
\be
{\cal L}=l^{-4}\left[1-\sqrt{1+l^4[(\d_\mu\phi)^2+m^2\phi^2]}\right].\label{dbi}
\ee
The reasoning is that in the high energy limit, many pions (lowest mass particles) will be created, so we need to consider a pion field as the
effective one, but this process is both nonperturbative and high energy, hence one needs a nonlinear action. As we will soon see, a polynomial interaction
does not have the required properties, so the DBI action is the natural one to consider.

In the high energy limit, colliding hadrons will look like pancakes due to Lorentz contraction, but moreover we need to consider them as just sources for
the pion field surrounding them, that will also get Lorentz contracted and look like a shockwave. Therefore the process considered in the asymptotic
regime is a collision of pion field shockwaves with the action (\ref{dbi}). We look for (classical) shockwave solutions to the action (\ref{dbi}).
The equations of motion are
\be
-\Box \phi+m^2\phi+l^4\frac{[(\d_\mu\d_\nu\phi)(\d_\mu\phi)\d_\nu \phi+(\d_\mu\phi)^2m^2\phi]}{1+l^4[(\d_\mu\phi)^2+m^2\phi^2]}=0.\label{dbieom}
\ee

The crucial simplification that allowed Heisenberg to do exact calculations is to consider that for a shockwave solution, only the physics near the shock
is relevant, and by focusing near that, we can ignore the dependence on the 2 transverse dimensions $y,z$ (with $r=\sqrt{y^2+z^2}$), and consider the
1+1 dimensional problem for time $t$ and longitudinal direction $x$ (along the direction of propagation).

Then from Lorentz invariance, he considers only solutions that depend on
\be\label{defstx}
s=t^2-x^2.
\ee
(Note that from now on, we will use $s$ to denote this variable only, and not the Mandelstam invariant, which will be called $\tilde s$)
This requires some explanation. The first point is that $\phi=\phi(s)$ is boost invariant for boosts in $x$:
Under a boost, we have $x^+\rightarrow e^\b x^+, x^-\rightarrow e^{-b}x^-$. But why do we need a boost invariant solution? The fact that $\phi$ is a scalar
means that $\phi'(x'^+,x'^-)=\phi(x^+,x^-)$, where $x^\pm =t\pm x$.
We could say that we find the solution $\phi(x^+,x^-)$ in a reference system and then define the one in another
reference system by $\phi'(x'^+,x'^-)=\phi(x^+,x^-)$, so any solution would work.

However, the essential point is that we use the ultra-relativistic approximation, in which even
though the pion is massive, we consider that the source moves on a lightcone, $x^+=0$ or $x^-=0$. As a result, we impose
that $\phi(x^+=0)=0$ or $\phi(x^-=0)=0$.
This in turn implies a power-law behaviour near the lightcone, i.e. (for $x^-=0$), $\phi\propto (x^-)^q$, $q>0$ for $x^-\sim 0$. But if we have an arbitrary
dependence on $x^+$, then in the boosted system $\phi'$ would have a power of $e^\b$ in front, unless we have the same power law for $x^+$, i.e. unless
$\phi(x^+,x^-)=\phi(x^+x^-)=\phi(s)$.

For $\phi=\phi(s)$, we have
\be
(\d_\mu\phi)^2=-4s\left(\frac{d\phi}{ds}\right)^2\;,
\ee
the DBI action becomes
\be
{\cal L}=l^{-4}\left[1-\sqrt{1+l^4\left(-4s\left(\frac{d\phi}{ds}\right)^2+m^2\phi^2\right)}\right]\;,
\ee
and its equation of motion becomes
\be
4\frac{d}{ds}\left(s\frac{d\phi}{ds}\right)+m^2\phi+\frac{8l^4s\left(\frac{d\phi}{ds}\right)^2}{1+l^4[-4s\left(\frac{d\phi}{ds}\right)^2+m^2\phi^2]}
\left[\frac{d\phi}{ds}-\frac{m^2\phi}{2}+2s\frac{d^2\phi}{ds^2}\right]=0.\label{question}
\ee
However, by multiplying with the denominator (assuming that it does not vanish), canceling and rewriting the terms we are led to the form
\be
4\frac{d}{ds}\left(s\frac{d\phi}{ds}\right)+m^2\phi=8sl^4\left(\frac{d\phi}{ds}\right)^2\frac{\left[\frac{d\phi}{ds}+m^2\phi\right]}
{1+l^4m^2\phi^2}.\label{heiseq}
\ee

When $m=0$, one can find an exact solution depending on an arbitrary parameter $a$,
\be\label{solm0}
\phi=\frac{1}{a}\log\left(1+\frac{a^2}{2l^4}s+\frac{a}{2l^4}\sqrt{4l^4s+a^2s^2}\right)\;,\;\;\;\; s\geq 0\;,
\ee
and $\phi=0$ for $s<0$.

When $m\neq 0$, one can find a perturbative solution at small $s$,
\be
\phi=\frac{\sqrt{s}}{l^2}(1+a\;s\; m^2+...)\;,\;\;\; 0\leq s\ll 1/m^2\;,\label{pertssol}
\ee
and $\phi=0$ for $s<0$, as well as a solution at large $s$,
\be
\phi\simeq \gamma s^{-1/4}m^{-1/2}\cos(m\sqrt{s}+\delta)\;,\;\;\; s\gg 1/m^2.
\ee

At this point, Heisenberg notes that for the model to be reasonable, we need that $(\d_\mu\phi)^2$ to be a finite constant at the position of the shock,
$s=0$, since we need the nonlinearities to play a role there. But for the free KG equation, the result is infinite, which is also unphysical.\footnote{The
KG equation for $\phi=\phi(s)$ is just $d/ds(s\;d\phi/ds)=0$, with the solution $\phi=A\log(s/s_0)$, which means $(\d_\mu\phi)^2=-4s(d\phi/ds)^2=-4A^2/s
\rightarrow \infty$.} Thus the only possibility to correctly describe the shock at $s=0$ is to have $(\d_\mu\phi)^2$ a finite constant, which leads to
$\phi\sim A\sqrt{s}$ for $s\rightarrow 0$. This, as we will show below,  is incompatible with an action with a canonical kinetic term and a polynomial potential.

\subsection{From the pion field to the nucleon-nucleon cross section}

The energy (Hamiltonian) density of the pion field is
\be
{\cal H}=\pi\dot\phi-{\cal L}=\frac{l^{-4}+(\nabla\phi)^2+m^2\phi^2}{\sqrt{1+l^4[(\d_\mu\phi)^2+m^2\phi^2]}}-l^{-4}\;,\label{Hamden}
\ee
and similarly the momentum density is
\be
{\cal P}=\pi\nabla\phi=\frac{ \dot\phi \nabla\phi }{\sqrt{1+l^4[(\d_\mu\phi)^2+m^2\phi^2]}}.\label{Momden}
\ee
Both densities have a denominator which is the square root term of the Lagrangian density.
In the massless case, substituting the solution (\ref{solm0}) into (\ref{Hamden})  we find that the energy density diverges at $s=0$
due to the denominator going to zero as $a\sqrt{s}/(2l^2)$. Similarly, in the massive case, substituting the solution (\ref{pertssol}) into (\ref{Hamden})
we find the same divergence due to the denominator going to zero at $s=0$ as $m\sqrt{s(1-6a)}$.

Now following  \cite{Heisenberg1952}
 we assume that one can introduce a small perturbation so that the denominator can be taken as a  non-vanishing constant.
In this case we can use the standard method of  Fourier transforming  (\ref{pertssol}) over $x$ to $k$, as
\be
\phi(k,t)=l^{-2}\int_0^t dx e^{ikx}\sqrt{t^2-x^2}(1+am^2(t^2-x^2)+...)\;,
\ee
which for $a=0$ (only the leading term) gives
\be
\phi(k,t)\simeq l^{-2}\frac{\pi}{2}\frac{|t|}{|k|}(J_1(|k||t|)+i\bf{H}_1(|k||t|))\;,\label{phikt}
\ee
where $J_1$ is a Bessel function and $\bf{H}_1$ is a Struve function.
When expanded at large $k$, we obtain
\be
\phi-l^{-2}i\frac{|t|}{|k|}\simeq \sqrt{-i}l^{-2}\sqrt{\frac{\pi}{2}}|t|^{1/2}|k|^{-3/2}e^{-i|k||t|}\left(1+\frac{3}{8|k||t|}e^{2i|k||t|}\right).\label{phiofkandt}
\ee
Note that the non-oscillatory part of $\phi$ is not a radiative piece, hence is dropped.

However, as discussed above,  in reality the shockwave should have a finite thickness in $\sqrt{s}$ of the order of the Lorentz contracted $1/m$, i.e.
$\sqrt{s}_{0m}\equiv \sqrt{s}_{min}=\sqrt{1-v^2}/m$, which means that at sufficiently large $t$, $\phi(k,t)$ should be cut off at
$k_{0m}=1/r_{0m}=\gamma m$, the
relativistic mass of the pion.
With the assumption of a constant denominator we get

\be
\frac{dE}{dk}\propto k^2\phi(k)^2\sim \frac{\rm const.}{k}\;,
\ee
where in the last equality we have substituted (\ref{phiofkandt}). But this is valid only for $k\leq k_{0m}$.

Finally, the momentum $k$ is
identified with the momentum of a pion $k_0$, and moreover the classical field close to the shock is identified with the classical limit of the
field of radiated pions in a hadron collision. Thus the radiated energy ${\cal E}$ (identified through canonical quantization with the pion field energy
$E$) per unit frequency of radiated pions is given by (denoting the constant by $B$)
\be
\frac{d{\cal E}}{dk_0}=\frac{B}{k_0}\;, \;\;\;\;
m\leq k\leq k_{0m}.
\ee
This integrates to
\be
{\cal E}=B\ln\frac{k_{0m}}{m}= B \ln\gamma\;,
\ee
and leads to a relation for the number of pions emitted for a given energy, since $dE=k_0 dn$, giving
\be
\frac{dn}{dk_0}=\frac{B}{k_0^2}\;,\;\;\; m\leq k_0\leq k_{0m}\;,
\ee
which integrates to
\be
n=\frac{B}{m}\left(1-\frac{m}{k_{0m}}\right).
\ee
Then the average emitted energy per pion is
\be
\langle k_0\rangle \equiv\frac{{\cal E}}{n}=m\frac{\ln(k_{0m}/m)}{1-m/k_{0m}}=m\frac{\ln \gamma}{1-\frac{1}{\gamma}}\simeq m\ln \gamma\;,
\ee
which is approximately constant (only logarithmic dependence on the energy).

The last step in the Heisenberg model is to assume that the emitted energy is proportional to the total energy of the system, $\sqrt{ \tilde s}$
(here $\tilde s$ is the Mandelstam variable), with the constant of proportionality (ratio of emitted energy) being approximately given by the
pion wavefunction overlap. Since at large transverse distance $r$ ($=\sqrt{y^2+z^2}$), the wavefunction is small $\phi l\ll 1$, thus it satisfies
the free massive KG equation, with solution $\phi(r)\sim e^{-mr}$, the wavefunction overlap is $\sim e^{-mb}$, where $b$ is the impact parameter, i.e.
transverse separation between the colliding hadrons at the impact point $x=0$. Then we have approximately
\be
{\cal E}\sim \sqrt{\tilde s}e^{-mb}.
\ee
The maximum impact parameter for which we have interaction, $b_{\rm max}$, arises when the emitted energy equals the average emitted energy per pion
$\langle k_0\rangle$, so that it corresponds to emitting just one pion. Then we have
\bea
\sqrt{\tilde s}e^{-mb_{\rm max}}&=&\langle k_0\rangle\Rightarrow b_{\rm max}=\frac{1}{m}\ln \frac{\sqrt{\tilde s}}{\langle k_0\rangle}\Rightarrow\cr
\sigma_{\rm tot}&=&\frac{\pi}{m^2}\ln^2\frac{\sqrt{\tilde s}}{\langle k_0\rangle}.
\eea
We see then that the saturation of the Froissart bound arises only if $\langle k_0\rangle$ is approximately constant as a function of energy.

Next we would like to compare this result with what one gets for an ``ordinary field theory"
  with a canonical kinetic term and a polynomial potential of the form
\be
{\cal L}=-\frac{1}{2}(\d_\mu\phi)^2-\frac{1}{2}m^2\phi^2-\lambda\phi^n\;,
\ee
The corresponding equation of motion resulting from it for the $\phi=\phi(s)$ ansatz,
\be
4\left(s\frac{d^2\phi}{ds^2}+\frac{d\phi}{ds}\right)+m^2\phi+n\lambda \phi^{n-1}=0\;,
\ee
do not have $\phi\sim A\sqrt{s}$ as a solution, since the first bracket is divergent, as it equals $-A/2\sqrt{s}$, and the other terms give zero, as they are
positive powers of $s$. In fact, we can see that the only way to satisfy the equation of motion at leading order in $s$ with a canonical kinetic term plus a
potential is for a potential that includes the logarithmic term $\Lambda \ln(\phi/\phi_0)$, since then in the equation of motion we have $\Lambda/\phi$,
and we can solve the equation with $A^2=2\Lambda$. But it is unclear how such a term could arise in the potential (especially since it is unbounded from
below at $\phi=0$).

Now  let's study  for this class of theories the energy per emitted pion.
In a way similar to the one described above we can prove that
\bea
\frac{d{\cal E}}{dk_0}&=&B\;,\;\;\; m\leq k_0\leq k_{0m};\Rightarrow \frac{dn}{dk_0}=\frac{B}{k_0}\;,\;\;\;\;
m\leq k_0\leq k_{0m}\Rightarrow\cr
\langle k_0\rangle &=&\frac{{\cal E}}{n}\simeq \frac{k_{0m}}{\ln\frac{k_{0m}}{m}}=m\gamma \frac{1}{\ln \gamma}\propto \frac{\sqrt{\tilde s}}{\ln
\sqrt{\tilde s}}.
\eea
That means that we don't get the saturation of the Froissart bound, but rather we get a constant $\sigma_{\rm tot}(\sqrt{\tilde s})$.

In fact, we can check that the saturation of the bound is obtained only for $d{\cal E}/dk_0\propto 1/k_0^n$, with $n\geq 1$, whereas for
actions with polynomial potentials this is not satisfied. In fact, as we saw,
$d{\cal E}/dk_0\propto 1/k_0$ was obtained from the behaviour $\phi\propto \sqrt{ s}$ of the
field near $s=0$, which was due to the DBI form of the action.

In conclusion, we have two physical ways to restrict the form of the action. As Heisenberg argued, we need $(\d_\mu\phi)^2$ to be a finite constant
in order to describe the correct physics, which restricts to $\phi(s)\propto \sqrt{s}$, arising only in DBI. On the other hand, if we are to be able to
saturate the Froissart bound (which should happen, as Froissart argued), we again need $d{\cal E}/dk_0\propto 1/k_0$, which again requires the DBI action.

\section{Uniqueness of the Heisenberg model action.}

In his paper \cite{Heisenberg1952}, Heisenberg shows that an action based on ordinary kinetic term with any kind of a potential term
does not saturate  the Froissart bound.
We now seek to check  how unique is the choice of Heisenberg of having the DBI action as the action of  the pion  field.
We have seen that we need an action with higher derivatives. The question is then   can we have other higher derivative actions?  In particular
we want to examine whether one needs an infinite series of  any power of the derivative term or it is enough to have certain finite series. And
furthermore if one needs an infinite series is the DBI action used by Heisenberg unique?

We now examine this question, by considering Lagrangeans of the type ${\cal L}(\phi, X)$, where $X=(\d_\mu \phi)^2$.

{\bf DBI truncated to first term.}

We will start by truncating the DBI Lagrangean from the previous section (with $m^2\phi^2$ promoted to $2V$ for more generality)
to the first interaction term, i.e.
\bea
{\cal L}&=&-\frac{1}{2}(\d_\mu\phi)^2-V(\phi)+\frac{l^4}{8}[(\d_\mu\phi)^2+2V(\phi)]^2\cr
&=&-\frac{1}{2}(\d_\mu\phi)^2-\tilde V(\phi)+\frac{l^4}{8}[(\d_\mu\phi)^2]^2+\frac{l^4}{2}(\d_\mu\phi)^2V(\phi)\;,
\eea
where $\tilde V(\phi)=V(\phi)-l^4V^2(\phi)/2$, but for generality we will consider arbitrary $\tilde V$.

Then the equation of motion is
\be
-\Box \phi+\tilde V'(\phi) +\frac{l^4}{2}(\d_\mu\phi)^2\Box\phi+l^4(\d_\mu\phi)(\d_\nu\phi)(\d_\mu\d_\nu\phi)+l^4(\d^2\phi)V(\phi)+\frac{l^4}{2}(\d_\mu\phi)^2
V'(\phi)=0\;,
\ee
and for $\phi=\phi(s)$ we get the equation of motion
\be
4\frac{d}{ds}\left(s\frac{d\phi}{ds}\right)(1-l^4V(\phi))+\tilde V'(\phi)+8sl^4\left(\frac{d\phi}{ds}\right)^2\left(2\frac{d\phi}{ds}+3s\frac{d^2\phi}{ds^2}
-\frac{V'(\phi)}{4}\right)=0.
\ee
We want to see whether a solution of the type $\phi=A\sqrt{s}$ near $s=0$ is possible. First we note that in this case, the terms with
$V$ and $V'$ are irrelevant (they are subleading),
so we will drop them for simplicity (we can add them for free at the end). Then, substituting, we get the equation of motion for the leading term
\be
\frac{A}{\sqrt{s}}\left(1+\frac{A^2l^4}{2}\right)=0\;,
\ee
so we see that for a real scalar field (as we want), when $A^2>0$, there is no solution.

{\bf Generalization with first derivative interaction}

Next, we drop the irrelevant $V$ terms and generalize by writing an arbitrary coefficient for the interaction term,
\be
{\cal L}=-\frac{1}{2}(\d_\mu\phi)^2+C[(\d_\mu\phi)^2]^2\;,
\ee
giving the equation of motion
\be
-\Box \phi+4C(\d_\mu\phi)^2\Box\phi+8C(\d_\mu\phi)(\d_\nu\phi)(\d_\mu\d_\nu\phi)=0\;,
\ee
and on $\phi=\phi(s)$, we get
\be
4\frac{d}{ds}\left(s\frac{d\phi}{ds}\right)+64Cs\left(\frac{d\phi}{ds}\right)^2\frac{d}{ds}\left(s\frac{d\phi}{ds}\right)
+64Cs\left(\frac{d\phi}{ds}\right)^2\left[\frac{d\phi}{ds}+2s\frac{d^2\phi}{ds^2}\right]=0.
\ee

Substituting the ansatz $\phi\simeq A\sqrt{s}$, we get
\be
\frac{A}{\sqrt{s}}(1+4CA^2)+16CA^2\left(\frac{A}{2\sqrt{s}}-\frac{A}{2\sqrt{s}}\right)=0.
\ee
Note that we have kept the last term, which is zero, for reasons to be explained later.

{\bf Generalization to arbitrary powers}

Next we consider on top of the previous, an arbitrary $n$-th order interaction,
\be
{\cal L}=-\frac{1}{2}(\d_\mu\phi)^2+C_2[(\d_\mu\phi)^2]^2+C_n[(\d_\mu\phi)^2]^n\;,
\ee
with equation of motion
\bea
&&-\Box \phi+4C_2(\d_\mu\phi)^2\Box\phi\left[1+\frac{nC_n}{2C_2}[(\d_\mu\phi)^2]^{n-2}\right]\cr
&&+8C_2(\d_\mu\phi)(\d_\nu\phi)(\d_\mu\d_\nu\phi)\left[1+\frac{n(n-1)C_n}{2C_2}[(\d_\mu\phi)^2]^{n-2}\right]=0\;,
\eea
and on the ansatz $\phi=\phi(s)=A\sqrt{s}$, we have
\bea
&&\frac{A}{\sqrt{s}}+4C_2A^2\frac{A}{\sqrt{s}}\left(1+\frac{nC_n}{2C_2}(-A^2)^{n-2}\right)\cr
&&+16C_2A^2\left(\frac{A}{2\sqrt{s}}-\frac{A}{2\sqrt{s}}\right)\left(1+\frac{n(n-1)C_n}{2C_2}(-A^2)^{n-2}\right)=0.
\eea

We can finally generalize to a sum of arbitrary powers,
\be
{\cal L}=-\frac{1}{2}(\d_\mu\phi)^2+\sum_{n\geq 2}C_n[(\d_\mu\phi)^2]^n\;,
\ee
with equation of motion
\be
-\Box \phi+\Box\phi\sum_{n\geq 2} 2nC_n[(\d_\mu\phi)^2]^{n-1}
+(\d_\mu\phi)(\d_\nu\phi)(\d_\mu\d_\nu\phi)\sum_{n\geq 2}4n(n-1)[(\d_\mu\phi)^2]^{n-1}=0.
\ee

On the solution $\phi=\phi(s)=A\sqrt{s}$, we get
\be
\frac{A}{\sqrt{s}}\left(1+\sum_{n\geq 2}2nC_n(-1)^nA^{2(n-1)}\right)+\left(\frac{A}{2\sqrt{s}}-\frac{A}{2\sqrt{s}}\right)\sum_{n\geq 2}8n(n-1)C_n(-1)^nA^{2(n-1)}
=0.\label{finalansatz}
\ee

For the DBI action, the coefficients, coming from the expansion of
\be
-(1+x)^{1/2}=-1-\frac{x}{2}-\frac{1/2(1/2-1)...(1/2-n+1)}{1\cdot 2\cdot ...\cdot (n)}x^n\;,
\ee
give therefore $sgn(C_n)=(-1)^n$, meaning that the coefficients inside the two brackets in (\ref{finalansatz}) are all positive.
Moreover, from the arguments in \cite{Adams:2006sv}, the signs coming from the DBI action are the ones needed for causality and locality
of an action (note that the metric convention in there is mostly minus, so all coefficients there are positive).

That means that for a general action, at any finite order in the terms, $\phi=A\sqrt{s}$ is not a solution.

But then the question is, how is it possible that the DBI action has this as a solution?
To answer that, we look at equation (\ref{question}), which is the equivalent of what we have here. On the ansatz
$\phi=A\sqrt{s}$, we get from it
\be
\frac{A}{\sqrt{s}}+\frac{2l^4A^2}{1-A^4l^2}\left(\frac{A}{2\sqrt{s}}-\frac{A}{2\sqrt{s}}\right)=0\;,
\ee
which at first seems not to have a solution, just like our finite order truncations, but looking better we see that $A^2=l^{-4}$ is a solution,
since then the second term is $0/0$, and there is a solution, as seen by going to the form (\ref{heiseq}). The essential fact is the existence of the
factor
\be
\frac{1}{1-x}=1+x+x^2+...+x^n+...
\ee
for $x=1$, multiplying the $(A/2\sqrt{s}-A/2\sqrt{s})=0$ term, but not the nonzero term. Thus in the case of the finite truncation, we have
the ratio of the zero and nonzero terms being
\be
\frac{\sum_{n\geq 2}8n(n-1)C_n(-1)^nA^{2(n-1)}}{1+\sum_{n\geq 2}2nC_n(-1)^nA^{2(n-1)}}\rightarrow\infty\;,
\ee
which goes to infinity for an infinite number of terms, allowing the solution.

In conclusion, the DBI action is the unique one satisfying the physical requirement $\phi(s)\simeq A\sqrt{s}$ near $s=0$
(it could be that there are other derivative actions, with an infinite number of terms, and the same
signs as DBI for the coefficients, but it is unlikely), however we can add a potential inside or outside the square root without modifying the
result.

\section{Generalizations}

We first consider a simple generalization, instead of just a mass term inside the square root, a general potential $V$, with Lagrangean
\be
{\cal L}=l^{-4}\left[1-\sqrt{1+l^4[(\d_\mu\phi)^2+2V(\phi)]}\right].
\ee
Its equation of motion is
\be
-\Box \phi+\d_\phi V(\phi)+l^4\frac{[(\d_\mu\d_\nu\phi)(\d_\mu\phi)\d_\nu \phi+(\d_\mu\phi)^2\d_\phi V(\phi)]}{1+l^4[(\d_\mu\phi)^2+2V(\phi)]}=0\;,
\ee
and for a solution $\phi=\phi(s)$, we obtain (after the same manipulations as in the Heisenberg case)
\be
4\frac{d}{ds}\left(s\frac{d\phi}{ds}\right)+V'(\phi)=8sl^4\left(\frac{d\phi}{ds}\right)^2\frac{\left[\frac{d\phi}{ds}+V'(\phi)\right]}
{1+2l^4V(\phi)}.
\ee
It is easy to check that this equation has again the same small $s$ solution (\ref{pertssol}) for $a=0$, i.e. the leading term, since
the terms with $V$ in the equation of motion are actually subleading with respect to the others. This in turn leads to
the same analysis of Heisenberg, so this generalization is allowed.

We can also consider adding $V$ outside the square root,
\be
{\cal L}=l^{-4}\left[1-\sqrt{1+l^4[(\d_\mu\phi)^2]}\right]-V(\phi)\;,
\ee
and we can again check that the same thing happens: the solution $\phi(s)=l^{-2}\sqrt{s}+...$
is still valid, since again the terms with $V$ in the equation of motion are subleading on the solution.

\subsection{Several mesons and sigma model}

We can also consider $N$ scalar fields, corresponding to having several scalar mesons, $\phi^i$, $i=1,...,N$ and for generality
consider it in $d+1$ dimensions.
A generalized DBI model would be
\be
{\cal L}=l^{-(d+1)}\left[h(\phi^i)-f(\phi)\sqrt{1+l^{d+1}[g_{ij}(\phi^k)(\d_\mu\phi^i)(\d_\mu\phi^j)+ 2V(\phi^i)]}\right].\label{dbimetricv}
\ee
Then when $h(\phi^i)=g(\phi^i)=1$, for small fields we keep only the leading term in the expansion of the square root, and obtain the usual sigma
model with a potential,
\be\label{lcan}
{\cal L}_2\approx -\frac12 g_{ij}(\phi^k)(\d_\mu\phi^i)(\d_\mu\phi^j)- V(\phi^i).
\ee
The equations of motion of the action (\ref{dbimetricv}) are
\bea
&& l^{-(d+1)}\left[\pa_{\phi^i} h(\phi) -\pa_{\phi^i} f(\phi)\sqrt{1-l^{d+1}[g_{ij}(\phi^k)(\d_\mu\phi^i)(\d_\mu\phi^j)- 2V(\phi^i)]}\right]\times\cr
&&\times \sqrt{1-l^{d+1}[g_{ij}(\phi^k)(\d_\mu\phi^i)(\d_\mu\phi^j)- 2V(\phi^i)]} + \cr
&&  f(\phi)\left[ \frac12\pa^{\phi^i} g_{jk}(\phi)(\pa_\mu\phi^j)\d^\mu \phi^k-\pa_{\phi^i} V(\phi)
-\pa_\mu[ g_{ij}(\phi)\pa_\mu\phi^j]\right] - \pa_{\mu} f(\phi)g_{ij}(\phi)\pa_\mu\phi^j \cr
&& -\frac12 l^{(d+1)} \frac{f(\phi)g^{ij}(\phi)\pa_\mu\phi^j}{1-l^{d+1}[g_{ij}(\phi^k)(\d_\mu\phi^i)(\d_\mu\phi^j)- 2V(\phi^i)]}\times\cr
&&\times \left(\pa_\mu[g_{ij}(\phi)(\d_\mu\phi^i)(\d_\mu\phi^j)- 2V(\phi^i)]\right)=0.\cr
&&
\eea

To analyze this case, first note that Heisenberg already considered the case of several mesons with DBI action, but that meant that the sum was
outside the square root,
\be
{\cal L}=l^{-4}\sum_a\left[1-\sqrt{1+l^4[(\d_\mu\phi^a)^2+m_a^2\phi_a^2]}\right].
\ee
That case worked in the same way as for a single meson. We now consider the generalization with the sum inside the square root, and a sigma model
metric,
\be
{\cal L}=l^{-4}\left[1-\sqrt{1+l^4\left[\sum_{ab}G_{ab}(\phi_c)(\d_\mu\phi^a)(\d_\mu\phi^b)+\sum_a m_a^2\phi_a^2\right]}\right]\;,\label{sigmamodel}
\ee
where we can replace everywhere the mass terms with a general potential, since as we already saw that doesn't change anything.

Now in terms of the asymptotic value of the cross section (the Froissart behaviour), nothing changes, since the maximum cross section is
governed by the pion of smallest mass, that has the largest wavefunction at large distances, according to the mechanism reviewed below.
What does change is the value of the cross section at intermediate energies, where now we have cross sections for emissions of various scalar mesons.

The equations of motion coming from the action (\ref{sigmamodel}) are
\bea
&&-\left[4\frac{d}{ds}\left(G_{ab}(\phi)s\frac{d\phi^b}{ds}\right)+m_a^2\phi_a^2\right]\left[1+l^4\left(-4sG_{ef}\frac{d\phi^e}{ds}\frac{d\phi^f}{ds}+
m_e^2\phi_e^2\right)\right]\cr
&&-l^4m_e^2G_{ab}\left(-4s\frac{d\phi^b}{ds}\frac{d\phi^e}{ds}\right)\phi^e
-8sl^4G_{ab}\frac{d\phi^b}{ds}\frac{d}{ds}\left[sG_{ef}\frac{d\phi^e}{ds}\frac{d\phi^f}{ds}\right]=0\;.
\eea

The simplifications that occured when $G_{ab}=\delta_{ab}$ do not occur anymore. However, the fields $\phi^a$ will have in general also the interpretation
of some brane coordinates in the gravity dual descriptions of section 8. In the Heisenberg case, we had a single field, corresponding to a single
coordinate transverse to the brane, but in general we can have many.
Then the origin of coordinate, corresponding to the position of the brane, must be a stable point. Around it, we can expand the metric as
$G_{ab}=\delta_{ab}+{\cal O}(|\phi|)$, and write an ansatz for the fields as
\be
\phi^a=A^a\sqrt{s}\;,
\ee
for $s\rightarrow 0$. At $s=0$, the fields are at 0, i.e. the stable point (the "IR brane" or IR cut-off of the gravity dual), and with the metric expanded
as above, we can check that the ansatz is a solution of the equations of motion if
\be
\sum_a (A^a)^2=1/l^4.
\ee

In order to understand the asymptotic cross sections, we consider the behaviour of the wavefunctions for large transverse $r$.
The large $r$ behaviour of the cross section is governed by the lightest meson, the pion. Indeed, the wavefunctions go like $\phi^a\propto e^{-m_ar}$ at
$r\rightarrow\infty$. Therefore, if we are in the asymptotic regime for the pion, $\phi^\pi\propto e^{-m_\pi r}$, which implies also
$|\phi^a|\ll |\phi^\pi|\ll l^{-1}$, then from the equations of motion we can check that we also have $\phi^a\propto e^{-m_ar}$, which by the usual Heisenberg
argument implies that
\bea
&&e^{-m_\pi b_\pi}\sim\frac{\langle k_0\rangle}{\sqrt{s}}\Rightarrow \sigma_{\rm tot}\simeq \sigma_\pi\simeq \frac{\pi}{m_\pi^2}\ln^2\left(\frac{\sqrt{s}}
{\langle k_0\rangle}\right)\cr
&&e^{-m_a b_a}\sim\frac{\langle k_0\rangle}{\sqrt{s}}\Rightarrow
\sigma_a\simeq \frac{\pi}{m_a^2}\ln^2\left(\frac{\sqrt{s}}{\langle k_0\rangle}\right)\ll \sigma_{\rm tot}\;,
\eea
where $\sigma_a$ is the cross section for production of mesons $a$. Therefore in the asymptotic regime, all the $\sigma_a$ should behave like
Froissart saturation, with corresponding coefficients $\pi/m_a^2$.

\subsection{AdS case and curved space generalizations; solutions}

In section 8 we will discuss possible relations between Heisenberg model and a holographic description of the nucleon-nucleon scattering.
We have seen in the previous subsection  that a natural generalization of  Heisenberg's model includes  several scalars corresponding
to several mesons, a sigma model for them.
Here we discuss a particular example of such a generalization which   is a DBI sigma model  in AdS spacetime.
This arises also naturally in the context of gauge/gravity duality.

The DBI  action on the flat worldvolume in $d+1$ dimensions, i.e. for a D$d$-brane, takes the form
\be\label{DBI}
S_{DBI}= T_d\int d^{d+1}\sigma e^{-\tilde\phi} \sqrt{ - \det[\pa_\mu X^i\pa_\nu X^j g_{ij}(X) + 2\pi\alpha' F_{\mu\nu}]}\;,
\ee
where $T_d$ is the D$d$-brane tension, $\tilde\phi$ is the dilaton, $\sigma_\mu$ are the world volume coordinates, $X^i$ $i=1,...D$ are the
target space coordinates  and $g_{ij}(X^k)$ is the metric on that target space.
The $d+1$ dimensional  DBI action  describes in particular the physics of D$d$-branes.  The D-brane action in fact also include a CS term,
but for our purposes, the effect of that will be just to subtract $\int d^{d+1}\sigma T_d$ from the above action.

Imposing $d+1$ dimensional Lorentz invariance, switching off the gauge fields, writing $T_d=l^{-(d+1)}$, using  the static gauge
$\sigma_\mu = \delta_\mu^I X^I$ for $I=0,1,2,d$, and defining the vector $\vec \phi\equiv X^i/l^{(d+1)/2}\equiv iv^i$ with $i=d+1,...,D$,
the DBI action reduces to
\be
S_{Dd}= l^{-(d+1)}\int d^{d+1} x  e^{-\tilde\phi} \left [ \sqrt{-\det\left( \eta_{\mu\nu}\tilde g(\phi)+ l^{d+1}
\pa_\mu\phi^i\pa_\nu\phi^j g_{ij}(\phi) \right )} -1  \right ].\label{dbraneaction}
\ee
For the special case of the D$3$-brane moving in an $AdS_5\times S^5$ space (the space generated by a large number $N$ of other D$3$-branes),
this action is the "highly effective action" for the ${\cal N}=4$ SYM theory on D$3$-branes  written recently in \cite{Schwarz:2013wra}, which takes the form
\be\label{HesenAds}
S_{D3}\sim \int d^4 x  \phi^4 \left[\sqrt{-det\left( \eta_{\mu\nu} +
\frac{\pa_\mu\vec\phi\cdot\pa_\nu\vec\phi}{\phi^4} \right )} -1
\right]\;,
\ee
where  now the dilaton is a constant, the target
space coordinates  are   $\vec \phi\equiv v^i$ with $I=4,...,9$,
and the metric on the target space was taken to be
\be
ds^2 = R^2\left[\phi^2\eta_{IJ} d x^I dx^J + \frac{1}{\phi^2} d\vec \phi \cdot
d\vec \phi\right].
\ee

For the special case of a single scalar with $\tilde \phi=0, \tilde g=1$, using the identity
\be
-\det \left [ \eta_{\mu\nu} + g(\phi)  \pa_\mu\phi\pa_\nu\phi \right ]= 1+g(\phi)\pa_\mu\phi\pa^\mu\phi\;,
\ee
the action for a D-brane (\ref{dbraneaction}) reduces to the Heisenberg type action with a $g(\phi)$, or one-dimensional sigma model action
of the type (\ref{sigmamodel}).

For the case of several scalars, but with the metric trivialized around the position of the brane, i.e. $g_{ij}(\phi)\simeq g(\phi)\delta_{ij}$,
we have
\bea
-\det (\eta_{\mu\nu}+g(\phi)\d_\mu\phi^i\d_\nu\phi^i)&=&
-\frac{1}{d!}\epsilon^{\mu_1...\mu_d}\epsilon^{\nu_1...\nu_d}(\eta_{\mu_1\nu_1}+g(\phi)\d_{\mu_1}\phi^i
\d_{\nu_1}\phi^i)...\cr
&&...(\eta_{\mu_d\nu_d}+g(\phi)\d_{\mu_d}\phi^i\d_{\nu_d}\phi^i)\cr
&=&\frac{1}{d!}\left[\epsilon_{\mu_1...\mu_d}\epsilon^{\mu_1...\mu_d}+dg(\phi){\epsilon^\mu}_{\nu_2...\nu_d}\epsilon^{\nu\nu_2...\nu_d}
\d_\mu\phi^i\d_\nu\phi^i\right.\cr
&&\left.+...\right]\cr
&=&1+g(\phi)\d_\mu\phi^i\d^\mu\phi^i+g^2(\phi)(\d_{\mu_1}\phi^i\d_{\nu_1}\phi^i)(\d_{\mu_2}\phi^j\d_{\nu_2}\phi^j)\delta_{\mu_1\mu_2}^{\nu_1\nu_2}\cr
&&+g^3(\phi)(\d_{\mu_1}\phi^{i_1}\d_{\nu_1}\phi^{i_1})(\d_{\mu_2}\phi^{i_2}\d_{\nu_2}\phi^{i_2})(\d_{\mu_3}\phi^{i_3}\d_{\nu_3}\phi^{i_3})
\delta_{\mu_1\mu_2\mu_3}^{\nu_1\nu_2\nu_3}\cr
&&+...
\eea
Then {\em on the solution } $\phi=\phi(s)$, we have
\bea
&&2(\d_{\mu_1}\phi^i\d_{\nu_1}\phi^i)(\d_{\mu^2}\phi^j\d_{\nu_2}\phi^j)\delta_{\mu_1\mu_2}^{\nu_1\nu_2}\cr
&=&\d_{\mu_1}\phi^i\d^{\mu_1}\phi^i\d_{\mu^2}\phi^j\d^{\mu_2}\phi^j
-\d_{\mu_1}\phi^i\d^{\mu_1}\phi^j\d_{\mu^2}\phi^i\d^{\mu_2}\phi^j\cr
&=& 16 s^2\frac{d\phi^i}{ds}\frac{d\phi^i}{ds}\frac{d\phi^j}{ds}\frac{d\phi^j}{ds}-16^2\frac{d\phi^i}{ds}\frac{d\phi^j}{ds}\frac{d\phi^i}{ds}\frac{d\phi^j}{ds}
=0
\eea
and we can easily see that for the higher terms the same happens. Therefore {\em on the solution $\phi=\phi(s)$}, the presence of higher order
terms inside the
square root in the D-brane DBI action (\ref{dbraneaction}) is not relevant, and we have still a sigma model action like (\ref{sigmamodel}).
That means that the Heisenberg analysis is still valid on the case of the general DBI D-brane action.

{\bf Shockwaves for D-brane in curved space}

Consider the action of a D3-brane moving in $AdS_5$, i.e. the "highly effective action" of (\ref{HesenAds}) for a single scalar $\phi$ and metric
$g(\phi)=\phi^{-4}$,
\be
{\cal L}=l^{-4}\left[1-\sqrt{1+\frac{(\d_\mu\phi)(\d_\mu\phi)}{\phi^4}
}\right].\label{dbiheis}
\ee
Its equation of motion on the ansatz $\phi=\phi(s)$ is
\be
s \phi'' + \phi' - 2 \frac{s}{\phi} (\phi')^2
- 2\frac{s}{\phi^4} (\phi')^3 =0 \;,
\ee
which is a special case of the more general form with an arbitrary metric $g(\phi)$,
\be\label{eqngeneral}
s \phi'' + \phi' + \frac12 \frac{g'(\phi)}{g(\phi)} {s}(\phi')^2 - 2s g(\phi) (\phi')^3 =0\;,
\ee
for $g(\phi) =\frac{1}{\phi^4}$.

It is easy to check that an exact solution for this non-linear
equation is
\be\label{solD3}
\phi(s) = \frac{1}{\sqrt{s}}=
\frac{1}{\sqrt{t^2-x^2}}.
\ee

Substituting  the solution into the Lagrangian density (\ref{dbiheis})
we find that the square root vanishes and ${\cal L} = l^{-4}$.
(The same holds for the solution of the massless Heisenberg model
where $g(\phi)=1$).

In fact one can use this property to find solutions for other target space metrics. For $ g(\phi) = l^{4-n}(\phi)^{-n}$, we
get
\be
{\cal L} = l^{-4}\ \ \ \rightarrow \qquad 4s\ g(\phi)
(\phi')^2=1\ \ \  \rightarrow \qquad \phi(s) = l^{-1} (l^{-2}s)^{\frac{1}{2-n }}\left[\frac{2-n}{4}\right]^{\frac{1}{2-n}}.\label{gennsol}
\ee

Furthermore, for a general metric $g(\phi)$ we find that the solution for $\phi(s)$ is
\be
\int d\phi \sqrt{g(\phi)} = \sqrt{s}.
\ee

It is easy to check that this solution solves indeed the equation
of motion (\ref{eqngeneral}).

We now want to see whether we can saturate the Froissart bound for $g(\phi)=\phi^{-4}$, with solution (\ref{solD3}).

Its Fourier transform is
\bea
\phi(t,k) &=& \frac{\pi}{2}(J_0(|k||t|)+{\bf L}_0(i|k||t|))\cr
&\rightarrow& \sqrt{\frac{\pi}{2}}\sqrt{-i}\frac{t^{-1/2}
e^{i| k| |t| }   \left(1+\frac{1}{8 i\left| k\right| |t|}\right)}{\sqrt{\left| k\right| }}\;,
\eea
where on the second line we have written the $k\rightarrow \infty$ limit, and $\bf{L}_0$ is a Struve function.

Then for the energy per pion frequency, we get
\be
\frac{d{\cal E}}{dk_0}=k^2|\phi(k,t)|^2\sim k.
\ee
As we already argued, in this case we do not get a saturation of the Froissart bound.

Similarly, for $g(\phi)=l^{4-n}\phi^{-n}$, the Fourier transform of the solution (\ref{gennsol}) gives at $\k\rightarrow \infty$,
\bea
&&\phi(t,k)-il^{-\frac{4-n}{2-n}}\left(\frac{2-n}{4}\right)^{\frac{1}{2-n}}\frac{t^{\frac{2}{2-n}}}{k}
\rightarrow l^{-\frac{4-n}{2-n}}(-i)^{\frac{3-n}{2-n}}(2-n)^{\frac{1}{2-n}}2^{-\frac{1}{2-n}}\Gamma\left(\frac{n-3}{n-2}\right)\times\cr
&&\times
t^{\frac{1}{2-n}} \left(\frac{1}{\left| k\right|}\right)^{\frac{n-3}{n-2}} e^{i\left| k\right| |t|} \left(1+\frac{n-3}{2(n-2)^2 i\left| k\right| |t|}\right).
\eea
This leads to
\be
\frac{d{\cal E}}{dk_0}=k^2|\phi(t,k)|^2\sim k^{\frac{2}{n-2}}\;,
\ee
which means we obtain saturation of the bound only for $n\in (0,2)$.

\subsection{Introducing vector mesons}

As we saw in (\ref{DBI}), the DBI action on the worldvolume of a D$d$-brane has vector fields. In fact, in AdS/QCD approaches with probe branes,
like for instance the Sakai-Sugimoto model or the models of section 8, these vectors on the gravitational side give rise on the dual field theory side
to towers of vector meson states. We are considering mainly a flat metric for the scalar fields (trivial sigma model), which arises as an approximation
in the IR of the gravity dual, as we discussed. Then we must consider the effect of the gravity dual metric (that drives the brane to the stable point
around which the metric is flat) to be to give masses to the fields.

Therefore we consider the DBI action with a mass for the vector inside the square root,
\bea
{\cal L}&=&l^{-4}\left[1-\sqrt{\det(\eta_{ab}+l^4\d_a\phi \d_b\phi+l^2F_{ab})+m^2\phi^2+M_V^2A_a^2}\right]\cr
&=&l^{-4}\left[1-\sqrt{1+l^4[(\d_\mu\phi)^2+m^2\phi^2]+\frac{l^4}{2}F_{ab}F^{ab}-l^8\left(\frac{1}{4}\tilde F_{ab}F^{ab}\right)^2+M_V^2A_a^2+...}\right]\cr
&&.
\eea
where $
\tilde F_{ab}=\frac{1}{2}\epsilon_{abcd}F^{cd}$ .

For the vector wavefunctions $A_a(r)$, like for the pion field $\phi(r)$, we need to give some initial data (boundary condition), and then the wavefunction
is determined from the equation of motion of the above action.

Let us consider first the case with no pions, just vector mesons, i.e. $\phi=0$. At sufficiently large $r$ we have again the usual free field decay
\be
A_a(r)=A_a e^{-M_Vr}\;,
\ee
and again, with the {\em additional} assumption that $\sigma_V$, the cross section for emission of $V$ vector mesons,
is obtained when the emitted vector meson energy (which we should calculate) equals the average per vector meson emitted energy, i.e.
\be
\frac{\langle k_0\rangle}{\sqrt{s}}=e^{-M_Vb_{max}}\;,
\ee
we obtain
\be
\sigma_V=\pi b_{max}^2.
\ee

Of course, the correct calculation would be the one where we have {\em both} the pions and the vector meson wavefunctions, and then we can calculate
$\sigma_V$ as above.

The action for only vector mesons with mass $M_V$ and no pions is
\be
{\cal L}=l^{-4}\left[1-\sqrt{1+\frac{l^4}{2}F_{ab}F^{ab}-l^8\left(\frac{\tilde F_{ab}F^{ab}}{4}\right)^2+l^4M_V^2A_a^2}\right]\;,
\ee

We would like to restrict again the dependence of the gauge fields to a dependence on $(t,x)$  and furthermore to only $s$ dependence,
but now with all the four vector  fields $A_a$, with $a=0,1,2,3$,
since  there is no  gauge invariance due to the fact that  the vector mesons are   massive ones.

Substituting $A_a=A_a(s)$ in $F_{ab}$, we find
\bea
F_{ab}F^{ab}&=&2[-(F_{01})^2-(F_{02})^2-(F_{03})^2+F_{12}^2+F_{13}^2]\cr
&=&-8[s\left(\frac{dA_2}{ds}\right)^2+s\left(\frac{dA_3}{ds}\right)^2+\left(t\frac{dA_1}{ds}+x\frac{dA_0}{ds}\right)^2\cr
\tilde F_{ab}F^{ab}&=&\frac{1}{2}\epsilon^{abcd}F_{ab}F_{cd}=\frac{1}{2}\epsilon^{0123}(8F_{01}F_{23}-8F_{02}F_{13}+8 F_{03}F_{12})\cr
&=& 8\left(-8t\frac{dA_2}{ds}x\frac{dA_3}{ds}+t\frac{dA_3}{ds}x\frac{dA_2}{ds}\right)=0\;,
\eea
so that the action for  the ansatz $A_a(s)$  is
\bea
{\cal L}&=&l^{-4}\left[1-\left(1-4l^4\left[s\left(\frac{dA_2}{ds}\right)^2+s\left(\frac{dA_3}{ds}\right)^2+\left(t\frac{dA_1}{ds}+x\frac{dA_0}{ds}\right)^2\right]
\right.\right.\cr
&&+l^4M_V^2(A_0^2+A_1^2+A_2^2+A_3^2)\Bigg)^{1/2}\Bigg].
\eea

Note that in this action we can consistently truncate $A_0=A_1=0$, and then the DBI action for $A_2$ and $A_3$ are the same as for
two DBI pions of Heisenberg,
for which we already saw that we need the full nonlinear DBI action.

\section{The pion wavefunction}

The pion wavefunction should be a solution of the equations of motion coming from the pion action. Following Heisenberg, we have considered only the
1+1 dimensional case of $\phi(s)$ that describes the physics near the shock, at $s\sim 0$, and the weak field case $\phi(r)$, spherically symmetric
in the transverse coordinates, so a function of only $r=\sqrt{y^2+z^2}$.

Note that in general, we do not need to have even an ansatz depending on both $r$ and $s$, i.e. $\phi(s,r)$, but rather depending
independently on all 4 coordinates, however considering $\phi(s,r)$ is a simple way to start the analysis.

\subsection{Possible generalizations to $\phi(r)$ and $\phi(s,r)$.}

{\bf Static spherically symmetric solutions.}

Consider first spherically symmetric solutions depending on all 3 coordinates, i.e. on $r=\sqrt{x^2+y^2+z^2}$. Moreover, generalize to $n$ space dimensions.
The Lagrangean (\ref{dbi}) becomes
\be
{\cal L}=l^{-4}r^{n-1}\left[1-\sqrt{1+l^4(\phi'^2+m^2\phi^2)}\right]\;,
\ee
and its equation of motion is
\be
\left(\phi''+\frac{n-1}{r}\phi'-m^2\phi\right)[1+l^4(\phi'^2+m^2\phi^2)]-l^4\phi'^2(\phi''+m^2\phi)=0.
\ee
where $'$ denotes differentiation  with respect to $r$.
After simplifications, it is rewritten as
\be
\phi''+\frac{n-1}{r}\phi'-m^2\phi=l^4\frac{\phi'^2}{1+l^4m^2\phi}\left(2m^2\phi-\frac{n-1}{r}\phi'\right).
\ee

{\em One dimensional solution.}

In $n=1$ space dimension, the ansatz
\be
\phi(r)=\frac{A}{1+\b r}
\ee
is an approximate solution.
Indeed upon substituting this ansatz into   the equation of motion, we obtain
\be
2\b^2-m^2(1+\b r)^2=\frac{2\b^2}{1+\frac{(1+\b r)^2}{l^4m^2A^2}}\;,
\ee
after simplifying by a common factor $A/(1+\b r)^3$. The ansatz satisfies the equation of motion, if $\b\gg m$ and
\be
\frac{2\b^2}{l^4m^2A^2}=m^2\;,
\ee
and then it is valid even in the $\b r\sim {\cal O}(1)$ regime, since then we can approximate
\be
\frac{2\b^2}{1+\frac{(1+\b r)^2}{l^4m^2A^2}}\simeq 2\b^2\left(1-\frac{(1+\b r)^2}{l^4m^2A^2}\right)\simeq 2\b^2-m^2(1+\b r)^2.
\ee
In conclusion, the solution is
\be
\phi(r)\simeq \frac{A}{1+\frac{Ar}{\sqrt{2}m^2l^2}}\;,
\ee
and as we can see, it is parametrized by $A$, and is valid for $A\gg m^3l^2$.

However, this solution is only valid in $n=1$ space dimension.

At very large distances, the wavefunction in any dimension becomes
\be
\phi(r)=Be^{-mr}\;,
\ee
which is what Heisenberg considered as well.

{\bf Solutions  of the form $\phi(s,r)$.}

A possible generalization that would include both the $\phi(s)$ near $s=0$ and the $\phi(r)$ near $r\rightarrow\infty$ is $\phi(s,r)$.
Substituting this ansatz in the DBI action, we obtain first
\be
(\d_\mu\phi)^2=-4s\left(\frac{d\phi}{ds}\right)^2+\phi'^2\;,
\ee
and then for the Lagrangean
\be
{\cal L}=l^{-4}r^{n-1}\left[1-\sqrt{1+l^4\left(\phi'^2-4s\left(\frac{d\phi}{ds}\right)^2+m^2\phi^2\right)}\right]\;,
\ee
where $n$ is now the number of transverse space dimensions ($n=2$ in the physical case).
As before, we find the equation of motion
\bea
&&\left(\phi''+\frac{n-1}{r}\phi'-m^2\phi-4\frac{d}{ds}\left[s\frac{d\phi}{ds}\right]\right)[1+l^4m^2\phi^2]\cr
&&+8l^4s\left(\frac{d\phi}{ds}\right)^2\left[\frac{d\phi}{ds}+m^2\phi\right]
-2m^2l^4\phi'^2\phi\cr
&&+\frac{n-1}{r}l^4\phi'\left[\phi'^2-4\left(s\frac{d\phi}{ds}\right)^2\right]\cr
&&-4sl^4\left(\frac{d\phi}{ds}\right)^2\phi''-4l^4\phi'^2\frac{d}{ds}\left[s\frac{d\phi}{ds}\right]+8l^4s\frac{d\phi}{ds}\phi'\frac{d\phi'}{ds}=0\;,\label{phisreq}
\eea
where the third line contains terms with mixed derivatives.

We can check that again at $s\simeq 0$, $\phi=A\sqrt{s}$ is a solution, but $\phi=A\sqrt{s}f(r)$ is {\em not} a solution at nonzero $r$,
since the leading terms in the equations of motion for such an ansatz are
\be
0\simeq -4\frac{d}{ds}\left[s\frac{d\phi}{ds}\right]+8l^4s\left(\frac{d\phi}{ds}\right)^3=-\frac{A}{\sqrt{s}}f(r)+\frac{A}{\sqrt{s}}A^2l^4f^3(r)\;,
\ee
and this equation  has as only solution $f(r)=\pm 1$. Therefore the solution at nonzero $r$ and $s\simeq 0$ must be of the type
\be
\phi\simeq A\sqrt{s} +s^n f(r)\;,\label{phiofsr}
\ee
where $n\geq 1$.

\subsection{Delta function shockwave?}

Before we continue with $\phi(s,r)$,
we want to address the issue of a possible delta function shockwave. In the gravity dual theory, the gravitational shockwaves
that scatter are delta function shockwaves \cite{Kang:2004jd,Kang:2005bj},  so one can ask whether the same happens  also in the  field theory picture.

We want then to try a delta function ansatz for a $\phi=\phi(x^-,r)$, where $x^-=(x-t)/\sqrt{2}$,
\be
\phi(x^-,r)=\delta(x^-)\Phi(r).
\ee

The equations of motion for $\phi(x^-,r)$ in $n=2$ transverse dimensions are
\be
\phi''+\frac{1}{r}\phi'-m^2\phi=l^4\frac{\phi'^2}{1+l^4m^2\phi}\left(2m^2\phi-\frac{1}{r}\phi'\right).\label{eomdelta}
\ee
Then for the delta function ansatz, in the denominator on the right hand side
of (\ref{eomdelta}) we have the 1 negligible with respect to the $\phi^2$ term (since the whole term is proportional
to a delta function, so the denominator is relevant only on the delta function, when the value is infinite), and the equation becomes (after simplifying the
common delta function on both sides)
\be
\Phi''+\frac{1}{r}\Phi'-m^2\Phi=2\frac{\Phi'^2}{\Phi}\left(1-\frac{1}{2m^2r}\frac{\Phi'}{\Phi}\right).
\ee
But we can easily verify that this equation has no solutions of the form $Ae^{-mr}$ at large distances, nor of $Ar^{-p}$ type, and if we put
$Ae^{-a\r^p}$ we find that the only solution is $p=2,\a=-m^2/2$, i.e. $Ae^{m^2r^2/2}$, which is clearly nonphysical.

We do have in fact the solution $Ae^{imr}$, but it is a complex solution for a real scalar, and $A\cos (mr)$ is not a solution (the equation is nonlinear,
so we do not have a superposition principle). So the conclusion seems to be that this case $\phi=\delta(x^-)\Phi(r)$ is unphysical.

In fact, there are some ways around that. We can consider a case
when the $\phi$ resembles much a delta function, but it has a finite thickness, and the height of the delta function is not only finite, but such that
the 1 in the denominator of (\ref{eomdelta}) actually dominates, so we get the equation of motion
\be
\Phi''+\frac{1}{r}\Phi'-m^2\Phi=l^4\Phi'^2\left(2m\Phi-\frac{1}{r}\Phi'\right).
\ee

Another possibility is to add by hand a source $\phi \delta(x^-)f(r)$ to the action, leading to the modified equation of motion
\be
\Phi''+\frac{1}{r}\Phi'-m^2\Phi=2\frac{\Phi'^2}{\Phi}\left(1-\frac{1}{2m^2r}\frac{\Phi'}{\Phi}\right)+l^2m\Phi\left(1+\frac{\Phi'^2}{m^2\Phi^2}\right)^{3/2}
f(r)\;,\label{sourceeq}
\ee
but we are still left with the issue of understanding the source-free shockwaves like Heisenberg's.

Instead, we can notice that we do not really need a delta function shockwave in $x^+$, only need that $T_{++}$ becomes infinite at $x^+=0$.
Normally
that happens because of a $\delta(x^+)$ in $T_{++}$ which implies also a $\delta(x^+)$ in the field (in the case of the gravity dual, delta function in the
metric). But in the case of the solution of Heisenberg, we just have an energy density that blows up slowly near $x^+=0$.

Indeed, near $s=0$, we have (see (\ref{Hamden}))
\be
{\cal H}\simeq \frac{\phi'^2}{m\sqrt{s}}\sim \frac{l^{-4}x^2}{ms^{3/2}}\;,
\ee
which blows up at $s=0$. Moreover, we can calculate $T_{++}$, which turns the $x^2$ in the numerator into $s$, implying
$T_{++}\simeq l^{-4}/m\sqrt{s}\rightarrow 0$. That means that there is a source at $s=0$, since $T_{++}$ becomes infinite there, i.e.
at $x^+=0$ and $x^-=0$ (two plane waves, travelling in opposite directions). In the next section we will study the source of the pion field in more detail.

\subsection{Perturbative solution near $r=0$}

We now return to $\phi(s,r)$ and consider the expansion near $r=0$ of $\phi(s,r)$. We have found the equation of motion (\ref{phisreq}), and the ansatz
(\ref{phiofsr}). We first plug this ansatz in the equation of motion for $n=1$, but we find that while at zeroth order we get zero for $A=l^{-2}$,
then at first order we do not have cancellation.

It means that we need to consider the next order in $\sqrt{s}$, namely $n=3/2$. Then we can check that the relevant terms are only
\bea
-4\frac{d}{ds}\left[s\frac{d}{ds}\right]&=&-\frac{A}{\sqrt{s}}-9f\sqrt{s}\cr
+8l^4s\left(\frac{d\phi}{ds}\right)^3&=&\frac{A^3l^4}{\sqrt{s}}+9A^2l^4f\sqrt{s}+...\;,
\eea
but now we see that with $A=l^{-2}$ we cancel both zeroth order and first order terms. Moreover, now we have other terms in the equation of motion contributing,
in particular
\bea
&&-m^2\phi-4\frac{d}{ds}\left[s\frac{d}{ds}\right]l^4m^2\phi^2+8l^4sm^2\phi \left(\frac{d\phi}{ds}\right)^2\cr
&=&
-m^2A\sqrt{s}-A^3l^4m^2\sqrt{s}+2A^3l^4m^2\sqrt{s}=0\;,
\eea
but these also cancel!

That means that we need to consider also the second subleading term in the expansion in $s$ of $\phi(s,r)$,
\be
\phi=A\sqrt{s}+s^{3/2}f(r)+s^{5/2}g(r)\;,
\ee
and check the terms of order $s^{3/2}$ in the equation of motion as well. Incidentally, we can check that considering a power $s^\a$ smaller than $5/2$
doesn't work either, since then again only the two terms above contribute to second subleading order, the first giving $-4\a^2g(r)s^{\a-1}$, and the
second giving $+6\a g(r) s^{\a-1}$, so they only cancel for $\a=3/2$, which is excluded (is the first subleading term).

Then we obtain
\bea
-4\frac{d}{ds}\left[s\frac{d}{ds}\right]&=&-\frac{A}{\sqrt{s}}-9f\sqrt{s}-25gs^{3/2}\cr
+8l^4s\left(\frac{d\phi}{ds}\right)^3&=&\frac{A^3l^4}{\sqrt{s}}+9A^2l^4f\sqrt{s}+27Al^4f^2s^{3/2}+15A^2l^4gs^{3/2}+...
\eea

The other terms on the first line of (\ref{phisreq}) give to order $s^{3/2}$
\be
s^{3/2}\left[f''+\frac{n-1}{r}f'-m^2f-m^4A+47m^2f\right]\;,
\ee
the terms on the second line do not contribute to this order, and the terms on the third line give $-A^2l^4f''s^{3/2}$. Summing up all the contributions,
and using the zeroth order condition $A=l^{-2}$, we obtain
\be
s^{3/2}\left[27l^2f^2-10g+\frac{n-1}{r}f'+46m^2f-m^4l^{-2}\right]\;,
\ee
and equating this to zero fixes $g(r)$ to be
\be
g(r)=\frac{1}{10}\left[27l^2f^2(r)+\frac{n-1}{r}f'(r)+46m^2f(r)-m^4l^{-2}\right].
\ee

The interpretation is that we can specify arbitrarily the   function  $f(r)$, or in another way specify the function
\be
\left.\left[s^{-3/2}\frac{d\phi}{dr}\right]\right|_{s=0}\;,
\ee
which is an initial data on the Cauchy surface $s=0$. Once this is given, the rest of the function $\phi$
should be fixed by the equation of motion.

\subsection{Perturbative solution near $r=\infty$}

However, we are interested instead on the behaviour at large, but finite $r$, needed for the calculation of $\sigma_{\rm tot}(s)$ through $b_{\rm max}$.

We know that at small field and derivatives, the DBI action reduces to the free massive scalar action. Indeed, viewed as an expansion in $l^4$,
or in nonlinearities of the field, the equations of motion reduce to zeroth order to the free equation
\be
\phi''+\frac{n-1}{r}\phi'-m^2\phi-4\frac{d}{ds}\left[s\frac{d\phi}{ds}\right]=0\;,
\ee
and so, under the assumption that the $s$ dependence is subleading, and we can ignore the last term involving only $d/ds$, we obtain at $mr\gg 1$
the solution
\be
\phi\simeq Ae^{-mr}\;,
\ee
as expected. Note that the $\phi'$ term in the equations of motion is subleading in $mr$ and it doesn't contribute to this order.

But we can be more precise, since the exact solution to the free equation of motion is known.
If we had $n=3$, the exact solution would be the Yukawa potential,
\be
\phi(r)=\frac{Ae^{-mr}}{r}.
\ee

For $n=2$, the exact solution is a bit more complicated. We can write the equation of motion at nonzero mass $m$ as
\be
\frac{d^2\phi}{d(imr)^2}+\frac{1}{imr}\frac{d}{d(imr)}\phi +\phi=0\;,
\ee
which matches the defining differential equations of the Bessel functions at index $\nu=0$,
\be
\frac{d^2Z_0}{dz^2}+\frac{1}{z}\frac{dZ_0}{dz}+Z_0=0\;,
\ee
and therefore we have
\be
\phi=Z_0(imr).
\ee

We want to choose the Bessel function of imaginary argument that decays exponentially at infinity. This is
\be
K_0(mr)=\frac{\pi i}{2}H_0^{(1)}(imr)\;,
\ee
giving for the scalar
\be
\phi(r)=AK_0(mr).
\ee
The asymptotics at $mr\rightarrow\infty$ give
\be
\phi(r)\simeq A\sqrt{\frac{\pi}{2mr}}e^{-mr}\;,
\ee
but we should also note the asymptotics at $mr\rightarrow 0$, where
\be
K_0(z)\simeq -\ln \frac{z}{2}I_0(z)\simeq -\ln \frac{z}{2}.\label{phifreeapprox}
\ee

To find corrections to this free solution,
we could think of expanding the equations of motion in $l^4$, or equivalently the mass dimension of the remaining expression (once $l^4$ is
removed), but besides the free terms above, all the other terms are linear in $l^4$, that is, of mass dimension 7 with respect to the rest.

We can instead take an ansatz that $\phi$  depends only  on $r$, and not on $s$, with a coefficient that is of the order of $l^4$, i.e.
\be
\phi=AK_0(mr)+Bg(r)\;,
\ee
and $B\propto l^4$. Then the full equation of motion reduces to
\be
B\left(g''(r)+\frac{1}{r}g'(r)-m^2 g(r)\right)=l^4\phi'^2\left[2m^2\phi-\frac{1}{r}\phi'\right].
\ee

With the assumption that $B\propto l^4$ we can consider on the right hand side only the order zero term with $A$, to obtain
\bea
&&B\left(g''(r)+\frac{1}{r}g'(r)-m^2 g(r)\right)\cr
&=&(ml)^4A^3\left(\frac{d}{d(mr)}K_0(mr)\right)^2\left[2K_0(mr)-\frac{1}{mr}\frac{d}{d(mr)}K_0(mr)\right].
\eea

However, even for that, we can only find the leading order solution at $mr\rightarrow \infty$. Then $g(r)\simeq e^{-3mr}/(mr)^{3/2}$
solves the equation to leading order, and we find the solution
\be
\phi\simeq A K_0(mr)  +\left(\frac{\pi}{2}\right)^{3/2}\frac{m^2l^4A^3}{4}\frac{e^{-3mr}}{(mr)^{3/2}}\label{phiofrapprox}
\ee
at $mr\rightarrow \infty$.

\section{The source for the pion field}

In this section we would like to understand what  is the source of the pion field, which is supposed to represent the nucleons.
To do so, we first look at the original Born-Infeld action for nonlinear electrodynamics, in terms of a field strength $F_{\mu\nu}$,
and apply the lessons learned to our DBI case, first for a static solution, then for the shockwave.

\subsection{The vector Born-Infeld case}

In the original paper of Born and Infeld on nonlinear electrodynamics \cite{Born:1934gh}, the issue of the source for solutions of the nonlinear Maxwell field
was explored, and in fact it was the crucial motivation for the work: to obtain a smooth "electron" solution to the equations of motion, free of singularities.

The BI Lagrangean can be written as
\be
{\cal L}=\sqrt{1+F-G^2}-1\;,
\ee
where we defined
\be
F\equiv\frac{1}{b^2}(\vec{B}^2-\vec{E}^2);\;\;\;\;\;
G\equiv\frac{1}{b^2}(\vec{B}\cdot \vec{E}).
\ee
Using these definitions  we can define quantities analogous to the quantities defined for electromagnetism in a medium, namely
\bea
\vec{H}&\equiv & b^2\frac{\d {\cal L}}{\d \vec{B}}=\frac{\vec{B}-G\vec{E}}{\sqrt{1+F-G^2}}\cr
\vec{D}&\equiv & b^2\frac{\d {\cal L}}{\d \vec{E}}=\frac{\vec{E}-G\vec{B}}{\sqrt{1+F-G^2}}.
\eea
The equations of motion and Bianchi identities   of the BI Lagrangean, that correspond to Maxwell's equations of the linear theory, are
\bea
&&\vec{\nabla}\times \vec{E}+\d_0\vec{B}=0;\;\;\;\; \vec{\nabla}\cdot \vec{B}=0\cr
&&\vec{\nabla}\times \vec{H}-\d_0\vec{D}=0;\;\;\;\; \vec{\nabla}\cdot \vec{D}=0.
\eea

The Hamiltonian density can be written as
\be
{\cal H}=\sqrt{1+P-Q^2}-1\;,
\ee
where
\be
P=\frac{1}{b^2}(\vec{D}^2-\vec{H}^2);\;\;\;\;
Q=\frac{1}{b^2}(\vec{D}\cdot\vec{H})\;,
\ee

The inverse relations for the fields are obtained from the Hamiltonian as
\bea
\vec{B}&=&b^2\frac{\d{\cal H}}{\d\vec{H}}=\frac{\vec{H}+Q\vec{D}}{\sqrt{1+P-Q^2}}\cr
\vec{E}&=&b^2\frac{\d{\cal H}}{\d\vec{D}}=\frac{\vec{D}+Q\vec{H}}{\sqrt{1+P-Q^2}}.
\eea

At zero magnetic field, $\vec{B}=\vec{H}=0$, the equations of motion and Bianchi identities reduce to
\be
\vec{\nabla}\times \vec{E}=0;\;\;\;\; \vec{\nabla}\cdot\vec{D}=0\;,
\ee
and we also find $Q=0, P=\vec{D}^2/b^2$, $G=0, F=-\vec{E}^2/b^2$. Then $\vec{\nabla}\cdot \vec{D}=0$ reduces to
\be
\frac{d}{dr}(r^2D_r)=0\;,
\ee
which admits a non-trivial solution of the form
\be
D_e=\frac{e}{r^2}\;,
\ee
 the same as in Maxwell theory. More precisely, the solution is at $r\neq 0$, which means that we have actually
\be
\vec{\nabla}\cdot \vec{D}=4\pi e \delta^3(r)\;,
\ee
which gives the integral formula (from Gauss's law)
\be
4\pi e =\int_{\Sigma_r}D_r\sigma=\int_{\Sigma_r}d\vec{S}\cdot \vec{D}.
\ee

So from the point of view of $\vec{D}$ (the field in the medium in electromagnetism, where
$\vec{D}=\epsilon_0\vec{E}+\vec{P}$ and $\vec{\nabla}\cdot \vec{D}=0$ in the absence of external sources, and otherwise just includes the charges
external to the medium) the sources are point-like as in Maxwell
theory.

For a static system we have $\vec{E}=-\vec{\nabla}A_0$, where $A_0$ is the zero's component of the  gauge field vector potential,
and it  is related to $\vec{D}$ by
\be
\vec{D}=\frac{\vec{E}}{\sqrt{1-\frac{\vec{E}^2}{b^2}}}\Rightarrow \frac{e}{r^2}=D_r=\frac{E_r}{\sqrt{1-\frac{E_r^2}{b^2}}}=\frac{-A_0'(r)}{
\sqrt{1-\frac{A_0'^2}{b^2}}}\;,
\ee
which implies for the electric field $E$
\be
-E_r=A_0'=\pm \frac{e/r_0^2}{\sqrt{1+r^4/r_0^4}}\;,
\ee
where
\be
r_0=\sqrt{\frac{e}{b}}
\ee
is a radius related to the radius of the electron. We then also obtain the electric potential
\be
A_0(r)=\frac{e}{r_0}f\left(\frac{r}{r_0}\right)\;,
\ee
where
\be
f(x)=\int_x^\infty\frac{dy}{\sqrt{1+y^4}}\;,
\ee
and we obtain that $f(0)\simeq 1.8541$ and $A_0\simeq 1.8541 e/r_0$.

The finite maximum of the electric field $\vec{E}$ is obtained at $r=0$, and equals $e/r_0^2=b$, as expected, since ${\cal L}=\sqrt{1-\vec{E}^2/b^2}$.

As in electromagnetism, we can regard the total charge $Q$ in the material as either $\int dV \vec{\nabla}\cdot \vec{D}$ or $\int dV \vec{\nabla}\cdot
\vec{E}$ with the difference that the former expression counts only the outside charge introduced, whereas the latter expression counts all the charge,
including the polarization response of the material, which tends to spread out the charge density.

Also now, we can define $4\pi \rho=\vec{\nabla}\cdot \vec{E}$ and find after an easy calculation that
\be
\rho=\frac{e}{2\pi r_0^3}\frac{1}{(r/r_0)(1+(r/r_0)^4)^{3/2}}
\ee
and we see that this charge density is spread out, going as $1/r^7$ at $r\rightarrow \infty$, but only as $1/r$ at $r\rightarrow 0$. We can also
verify the fact that its integral gives the same result as the integral of $D_r$, namely $e$.

\subsection{The scalar DBI action and its source}

{\bf Static scalar DBI results}

A similar thing happens for the scalar DBI action. We start by reviewing the construction of the {\em static} scalar solutions paralleling the nonlinear
electrodynamics solutions, as presented in \cite{Nastase:2005pb}.

On static solutions, $\d_t\phi=0$, the scalar DBI action reduces to
\be
{\cal L}=\sqrt{1+\vec{F}^2}\;,
\ee
where
\be
\vec{F}\equiv \vec{\nabla}\phi\;,
\ee
$\phi$ being the DBI scalar. Note then that this action is the same as the vector BI action above for the case $\vec{B}=0$, just changing the sign inside the
square root. Therefore we can follow the same analysis, and first define
\be
\vec{C}=\frac{\d{\cal L}}{\d\vec{F}}=\frac{\vec{F}}{\sqrt{1+\vec{F}^2}}.
\ee
In terms of it, the equation of motion is
\be
\vec{\nabla}\cdot \vec{C}=0\;,
\ee
solved by
\be
C_r=\frac{e}{r^2}\;,
\ee
so that really we have $\vec{\nabla}\cdot \vec{C}=4\pi e$ in 3 spatial dimensions. Therefore the solution for the scalar is given by
\be
F_r=\d_r\phi =\frac{e/r^2}{\sqrt{1-e^2/r^4}}\;,
\ee
which is called the "catenoid". The solution has a horizon-like structure at $r=\sqrt{e}\equiv r_0$, due to the fact that it has the interpretation (in the case it is
the action of a D-brane) of one half of a D-brane-anti-D-brane solution connected by a throat.

{\bf The  DBI scalar shockwave-}

In the spirit of the model of \cite{Heisenberg1952} we now consider a four dimensional scalar field $\phi(r,s)$. In particular $\phi(s)$ can be recast from a 1+1 dimensional action  which  for the  massless case  reads
\be
{\cal L}=l^{-4}\left[1-\sqrt{1-4l^4s\left(\frac{d\phi}{ds}\right)^2}\right].
\ee

We define first the analog of the electric field from the Born Infeld paper,
\be
E_s=2\sqrt{s}\frac{d\phi}{ds}.
\ee
In terms of it, the Lagrangean becomes
\be
{\cal L}=l^{-4}\left[1-\sqrt{1-E_s^2}\right]\;,
\ee
just like the BI vector case. Then we also define the analog of the electric induction,
\be
D_s=\frac{\d{\cal L}}{\d E_s}\;,
\ee
which gives
\be
D_s=\frac{E_s}{\sqrt{1-l^4E_s^2}}.
\ee
The equation of motion (the analog of Maxwell's equation) is
\be
\frac{d}{ds}(\sqrt{s}D_s)=0\;,
\ee
solved by
\be
D_s=\frac{A}{\sqrt{s}},\;s> 0.
\ee
Causality then requires that we have $D_s=0$ for $s<0$. Then inverting $D_s(E_s)$ we get
\be
E_s=\frac{D_s}{\sqrt{1+l^4D_s^2}}=\frac{A/\sqrt{s}}{\sqrt{1+l^4A^2/s}}.
\ee
But since $D_s=0$ for $s<0$, we also have $E_s=0$ for $s<0$, which means that really,
\be
D_s=\frac{A}{\sqrt{s}}\theta(s);\;\;\;\;
E_s=\frac{A\theta(s)}{\sqrt{s+l^4A^2}}.
\ee

We can also integrate the above to find that $\phi$ is given by
\be
\phi=\int ds \frac{E_s}{2\sqrt{s}}=\frac{A}{2}\int ds \frac{1}{\sqrt{s(s+l^4A^2)}}=A\log\left[\frac{\sqrt{s}+\sqrt{s+l^4A^2}}{l^2A}\right]\;,
\ee
at $s>0$ and 0 at $s<0$ which has the same structure as of (\ref{solm0}).

This reduces at  small $s$ to
\be
\phi(s)\simeq l^{-2}\sqrt{s}\theta(s)\;,
\ee
which is the same solution as Heisenberg's. Note that the constant $A$ determining $D_s$ is arbitrary, even though $\phi(s)$ near $s=0$ is completely determined.

Then the electric field is a step function,
\be
E_s=2\sqrt{s}\frac{d\phi}{ds}\simeq l^{-2}\theta(s)\;,
\ee
and the electric induction is
\be
D_s=\frac{A}{\sqrt{s}}\theta(s).\label{induction}
\ee

Plugging back in the equation of motion for $D_s$, we have really for the analog of $\rho_{\rm ext.}=\vec{\nabla}\cdot \vec{D}$,
\be
\frac{d}{ds}(\sqrt{s}D_s)=\frac{d}{ds}(A\theta(s))=A\delta(s).
\ee

So as in the BI case, there is a source term, which is a delta function when viewed from the point of view of the induction $D_s$ (i.e., it is an
"external source" to the medium). The value of the charge, $A$, is arbitrary, even though $\phi(s)$ near $s=0$ is completely determined.

We can also define the equivalent of the $\vec{\nabla}\cdot \vec{E}=\rho$, the total charge (including the one due to the "polarization of the medium"),
which is spread out. We define the density
\be
\rho=\frac{d}{ds}(\sqrt{s}E_s)=A\frac{d}{ds}\left(\frac{\sqrt{s}\theta(s)}{\sqrt{s+l^4A^2}}\right)=\frac{l^4A^2}{2\sqrt{s}(s+Al^4A^2)^{3/2}}\theta(s).
\ee
Note that we dropped a term coming from the derivative of $\theta(s)$, proportional to $\sqrt{s}\delta(s)$, since this is zero. We see that this charge
drops at infinity as $1/s^2$, and at 0 only as $1/\sqrt{s}$, and integrates to the same total value as the one defined via $D_s$,
\be
\left.A\frac{\sqrt{s}\theta(s)}{\sqrt{s+l^4A^2}}\right|_0^\infty=A.
\ee

In conclusion, there is an "external source" located at $s=0$ (the shock's position), with an arbitrary charge, but the "in medium" source is spread out,
over an $s$ of the order of $l^4A^2$.

\section{The cross section}

We can now finally consider the calculation of cross sections arising from the Heisenberg model.

\subsection{Corrections away from the Froissart limit}

The first issue to address is of a systematic expansion away from the limit of Froissart bound saturation. It is clear that by considering a $\phi(r)$
that is not yet completely dominated by the $e^{-mr}$ term, we can find corrections to the Froissart behaviour of the cross section. If we have an exact
wavefunction, we can  obtain a $\sigma_{\rm tot}(\tilde s)$ that would be different in the leading behaviour, like a power law $\sigma_{\rm tot}(\tilde s)
\propto {\tilde s}^\a$, appearing before the onset of Froissart saturation.

{\bf Corrections to leading behaviour}

We first consider corrections to the $e^{-mr}$ behaviour of $\phi(r)$, which were found in (\ref{phiofrapprox}), with the free part being
asymptotically (\ref{phifreeapprox}). The $e^{-3mr}$ behaviour is subleading with respect to the $1/\sqrt{r}$ in the first factor, so we consider
\be
\phi(r)\propto \frac{e^{-mr}}{\sqrt{mr}}.
\ee
Then as usual,
the emitted energy is proportional to $\phi(b)\sqrt{\tilde s}$, and when it
gets down to $\langle k_{0,\pi}\rangle$ (the average per pion emitted energy), we reach $b_{\rm max}$. Thus
\be
\sqrt{\tilde s}\frac{e^{-m_\pi b_{\rm max}}}{\sqrt{b_{\rm max}m_\pi}}\simeq \langle k_{0,\pi}\rangle\;,
\ee
giving
\be
b_{\rm max}\simeq \frac{1}{m_\pi}\ln\frac{\sqrt{\tilde s}}{\langle E_\pi\rangle\sqrt{\ln( \sqrt{\tilde s}/\langle k_{0,\pi}\rangle)}}
\simeq \frac{1}{m_\pi}\ln\frac{\sqrt{\tilde s}}{\langle k_{0,\pi}\rangle}-\frac{1}{2m_\pi}
\ln \left[\ln\frac{\sqrt{\tilde s}}{\langle k_{0,\pi}\rangle}\right]\;,
\ee
and $\sigma_{\rm tot}(\tilde s)=\pi b_{\rm max}(\tilde s)^2$.

{\bf Possible new regime}

But besides the small corrections to the Froissart saturation regime above, we can in principle have also a situation where a new regime for
$\sigma_{\rm tot}(\tilde s)$ appears.

To avoid the leading Froissart behaviour, we must avoid the exponential $e^{-mr}$ for $r=b_{\rm max}$, so we need to have $m_\pi b_{\rm max}(\tilde s)<1$.
This can indeed exist in some energy regime $\tilde s$, for small mass $m=m_\pi\ll l^{-1}$.

Since the scale $l$ in Heisenberg's DBI action can presumably be identified with $\Lambda_{\rm QCD}$, and $\Lambda_{\rm QCD}\sim 2 m_\pi$,
the corrections of order $(ml)^2\sim (m_\pi/\Lambda_{\rm QCD})^2=1/4$ are small, so it could be a good approximation.

But if $ml\ll 1$, there is a regime where the wavefunction is linear, and when solving for $\phi(r)$ from the equation of motion we never get into
the nonlinear regime. That means that the full solution to the free equation, $\phi=A K_0(mr)$, is exact. At distances $r\ll m^{-1}$, we obtain
\be
\phi(r)\simeq -A\ln \frac{mr}{2}.
\ee

Then the condition for $b_{\rm max}$ at energies $\tilde s$ for which the above $\phi(r)$ are still in the linear regime is
\be
\sqrt{\tilde s}\left[-\frac{1}{2}\ln(m_\pi b_{\rm max}(\tilde s))\right]=\langle k_{0,\pi}\rangle\;,
\ee
giving
\bea
b_{\rm max}(\tilde s)&=&\frac{1}{m_\pi}e^{-2\frac{\langle k_{0,\pi}\rangle}{\sqrt{\tilde s}}}\Rightarrow\cr
\sigma_{\rm tot}(\tilde s)&=&\pi b_{\rm max}(\tilde s)^2=\frac{\pi}{m_\pi^2}e^{-4\frac{\langle k_{0,\pi}\rangle}{\sqrt{\tilde s}}}\;,
\eea
for $\sqrt{\tilde s}> \langle k_{0,\pi}\rangle$, which gives a mildly increasing dependence, that could be
easily mistaken for a small power law or the $\log^2$ behaviour of Froissart saturation.

In conclusion, such a new energy regime could appear in QCD just before the onset of Froissart saturation,
but it would be hard to distinguish experimentally from the small power law ("soft Pomeron")  behaviour, or from the Froissart saturation behaviour.

\subsection{Black disk model and ratio of elastic to total cross section}

Until now we have discussed the total cross section, or in the case of several mesons, also individual meson cross sections.
But we want now to  discuss also  the elastic cross section. For that however, we need a quantum amplitude, whose forward part gives the
total cross section, and whose absolute value squared gives the elastic cross section.

Since we do not have a quantum amplitude, only a total cross section, we can engineer an amplitude that gives this total cross section, and from
it calculate the elastic amplitude. The simplest model is  a black disk eikonal amplitude, with S matrix $S=e^{i\delta}$ and
Im$(\delta)=\infty$ for $b\leq b_{\rm max}(\tilde s)$ and with $\delta=0$ for $b> b_{\rm max}(\tilde s)$. This reproduces the
cross section $\pi b_{\rm max}(\tilde s)^2$.

For massless states scattering, we have in general
\bea
\frac{1}{\tilde s}{\cal A}(\tilde s,t)&=&-i\int d^2b e^{i\vec{q}\cdot \vec{b}}\left(e^{i\delta(b,\tilde s)}-1\right)\cr
&=&i\int_0^{b_{\rm max}(\tilde s)}bdb\int_0^{2\pi}d\theta e^{iqb\cos \theta}\left(e^{i\delta}-1\right)\;,\label{blackamp}
\eea
where $\vec{b}$ is the impact parameter (transversal), and its Fourier conjugate is $\vec{q}$, with $\vec{q}^2=t$.

For the black disk eikonal,
\be
\frac{1}{\tilde s}{\cal A}(\tilde s,t)=2\pi i \frac{b_{\rm max}(\tilde s)}{\sqrt{t}}J_1(\sqrt{t}b_{\rm max}(\tilde s)).
\ee

The total cross section is found from
\be
\frac{1}{\tilde s}{\rm Im}{\cal A}_{\rm elastic}(\tilde s, t=0)=\sigma_{\rm tot}(k_1,k_2\rightarrow {\rm anything})\;,
\ee
and it is easy to calculate that for the black disk eikonal we get $\sigma_{\rm tot}=\pi b_{\rm max}(\tilde s)^2$.

We should note here that most of the times, like for instance in \cite{Block:1984ru}, the black disk eikonal model starts with a partial amplitude
$a_l(k)=(e^{2i\delta_l(k)}-1)/(2i)$, suggested by the partial wave expansion,
which is a factor of 2 smaller than (\ref{blackamp}). After the normalization of the cross section is properly taken
into account, this leads to $\sigma_{\rm tot}=2\pi b_{\rm max}^2$ and, since $\sigma_{\rm tot}\sim {\rm Im} a$, but $\sigma_{\rm el}\sim |a|^2$,
so a rescaling of $a$ leads to a rescaling of $\sigma_{\rm el}/\sigma_{\rm tot}$, to an
$\sigma_{el}=\pi b_{\rm max}^2$. But our model, also used for instance in \cite{Kang:2004jd}, is physically different, since we considered simply, as
usual, the amplitude as the Fourier transform of the T-matrix, and $S=1+iT=e^{i\delta}$. This leads to $\sigma_{\rm tot}=\pi b_{\rm max}^2(\tilde s)$, which
we believe is a model more deserving of the name black disk, as the total cross section equals the classical one. Then, as we shall see, we obtain
$\sigma_{\rm el}/\sigma_{\rm tot}\simeq 1/4$, instead of 1/2.

In the case that the particles are massive with mass $m$ instead, the $1/\tilde s$ is replaced by $1/(2p_{CM}E_{CM})$. But if $m_1=m_2=m$,
$E_{CM}=2\sqrt{p_{CM}^2+m^2}=\sqrt{\tilde s}$, so we have
\be
2E_{CM}p_{CM}=\sqrt{\tilde s(\tilde s-4m^2)}.
\ee

On the other hand, for the differential cross section, we have the center of mass formula
\be
\left.\frac{d\sigma_{el}}{d\Omega}\right|_{CM}=\frac{|{\cal A}|^2}{64\pi^2E_{CM}^2}\;,
\ee
and the relativistically invariant differential cross section is
\be
\frac{d\sigma}{dt}=\frac{|{\cal A}(\tilde s,t)|^2}{16\pi \tilde s(\tilde s-4m^2)}.
\ee

For the black disk eikonal, we obtain
\be
\sigma_{\rm elastic}=\frac{4\pi^2b_{\rm max}^2(\tilde s)\tilde s(\tilde s-4m^2)}{16\pi \tilde s(\tilde s-4m^2)}\int \frac{dt}{t}[J_1(\sqrt{t}b_{\rm max}(\tilde s))]^2\;,
\ee
and since $\sigma_{\rm tot}=\pi b_{\rm max}^2(\tilde s)$, get
\be
\frac{\sigma_{\rm elastic}}{\sigma_{\rm tot}}=\frac{1}{4}\int \frac{dt}{t}[J_1(\sqrt{t}b_{\rm max}(s))]^2.
\ee

It remains to define the range of integration for $t$, given $\tilde s$. In the center of mass system,
\be
\tilde s=E_{CM}^2;\;\;\;\;
t=(\vec{p}_{CM}-\vec{k}_{CM})^2=k_{CM}^2+p_{CM}^2-2k_{CM}p_{CM}\cos\theta\;,
\ee
where $\vec{p}_{CM}$ and $\vec{k}_{CM}$ are momenta of the same particle, before and after the collision in the center of mass.
Then the range of integration for $t$, given $\tilde s$, which fixes $p_{CM}$ and $k_{CM}$, is
\be
t\in [(p_{CM}-k_{CM})^2,(p_{CM}+k_{CM})^2].
\ee

But $p_{CM}=k_{CM}$ and $\sqrt{\tilde s}/2=E_{CM}/2=E=\sqrt{p_{CM}^2+m^2}$, meaning that
\be
p_{CM}=k_{CM}=\sqrt{\frac{\tilde s}{4}-m^2}\;,
\ee
and then the range of integration of $t$ is
\be
t\in [0,\tilde s-4m^2]\;,
\ee
so that finally
\be
\frac{\sigma_{\rm elastic}}{\sigma_{\rm tot}}=\frac{1}{4}\int_0^{\tilde s-4m^2} \frac{dt}{t}[J_1(\sqrt{t}b_{\rm max}(\tilde s))]^2.
\ee

By using the recurrence relations for $J_\nu(x)$, we do the integral and obtain
\be
\frac{\sigma_{\rm elastic}}{\sigma_{\rm tot}}=\frac{1}{4}\left[1-\left(J_0(|b_{\rm max}(\tilde s)|\sqrt{\tilde s-4m^2})\right)^2
-\left(J_1(|b_{\rm max}(\tilde s)|\sqrt{\tilde s-4m^2})\right)^2\right].
\ee

At large $z$,
\bea
J_0(z)&\simeq &\sqrt{\frac{2}{\pi z}}\cos (z-\pi/4)\cr
J_1(z)&\simeq &\sqrt{\frac{2}{\pi z}}\cos (z-3\pi/4)= \sqrt{\frac{2}{\pi z}}\sin (z-\pi/4)\;,
\eea
so that finally we obtain
\be
\frac{\sigma_{\rm elastic}}{\sigma_{\rm tot}}\simeq \frac{1}{4}\left[1-\frac{2}{\pi b_{\rm max}(\tilde s)\sqrt{\tilde s-4m_N^2}}\right]
\ee
where we put $m_N$ for a nucleon or nucleus mass, corresponding to the case when we collide nucleons or nuclei. Then from the Heisenberg model $b_{\rm max}(s)\simeq
1/m_\pi \ln(s/s_0)$, so that the sought-for ratio is
\be
\frac{\sigma_{\rm elastic}}{\sigma_{\rm tot}}\simeq \frac{1}{4}\left[1-\frac{2m_\pi}{\pi \ln(\tilde s/s_0)\sqrt{\tilde s-4m_N^2}}\right]\;,
\ee
asymptoting very fast to 1/4.

This compares very well with the experimental results from the TOTEM experiment \cite{Csorgo:2012dm}.

\section { Heisenberg model and Holography}

In section 3.3 we described a sigma model in $AdS$ space. This can be directly related to another holographic model, the ``hard wall" model,
which is an $AdS$ background  chopped off at a certain value of the radial coordinate. This scenario is addresssed in the following subsection.
We then present an alternative approach that includes a  systematical analysis of  the relations  between
 Heisenberg's model and  the holographic description of a proton-proton scattering. This in fact  involves two steps. In the first we will
 establish the relations between the DBI action used in Heisenberg's model and the DBI action that emerges as the action of flavor
 branes in confining backgrounds.
The second step is to layout the holographic dual of scattering of baryons and to relate it to the extraction of the cross section from
Heisenberg's model. The two steps are described in the second and third subsections of this section.

\subsection{The relation to the holographic ``hard wall" model}

The remarkable fact is that, even though the Heisenberg's model was proposed before string theory was discovered,
the DBI action used by Heisenberg emerges  naturally in holographic models of QCD since it relates to the effective action of open strings.
In the simplest model for high energy QCD scattering introduced by Polchinski and Strassler, one
considers an $AdS_5$ space,
\be
ds^2=\frac{r^2}{R^2}d\vec{x}^2+R^2\frac{dr^2}{r^2}\;,
\ee
cut off at an $r_{\rm min}=R^2\Lambda$, with $\Lambda$ identified with the (pure) QCD scale (glueball scale). It was soon realized that one can think
of the IR cut-off as a dynamical IR brane (like in the Randall-Sundrum model), and the appropriately normalized fluctuation in the position $r_{\rm min}$,
the scalar $\sim \phi$ can be identified with the pion in QCD. But the action for the fluctuation in position of a brane is exactly the DBI action!

The only nontrivial part of the action is the potential for the brane position, which can appear, depending on the mechanism, either inside or outside
the square root.

The picture of high energy scattering is also similar in the gravity dual
\cite{Kang:2004jd,Kang:2005bj,Nastase:2005rp,Nastase:2006eb,Freund:2008tv,Nastase:2008hw}.
In a purely gravitational theory, we have gravitational shockwave collision,
happening near the IR cut-off, creating a black hole on the IR cut-off, being mapped to the pion field shockwave collisions creating a fireball.
But more precisely, when we consider also the fluctuation of the IR cut-off giving the pion, we have the same picture, of pion field shockwaves colliding
and creating a fireball.

\subsection{The DBI action of flavor branes in confining backgrounds}

Heisenberg's model assumes that the scalar fields that are in charge of the interaction between nucleons are governed by a DBI action in
flat space-time.  Holography provides dual string descriptions to certain strongly coupled gauge dynamical systems.   As was mentioned
above, the DBI action is a basic tool in the toolkit of string theories. Thus, an obvious question to ask is whether  one can relate Heisenberg's
model to a holographic description of proton-proton scattering, and  in what way. To answer this question one has to address first the issue of
what is the holographic laboratory dual of QCD in its confining phase.

The basic $AdS_5\times S^5$ string theory, the dual of ${\cal N}=4$ SYM is clearly not the right setup. It is both conformal and maximally
supersymmetric. One has to deform the geometrical background in such a way that the isometry group is not $SO(4,2)\times SO(6)$ but
rather only the four dimensional Poincar\'{e} symmetry. Obviously the desired background should be
equipped with a scale which  breaks  scale invariance. To check whether a given background corresponds to a boundary confining field theory,
 one should investigate the stringy dual of the  Wilson line. A necessary condition for a ``confining background" is that   any  rectangular
 Wilson line along a space and  the time directions  should admit a confining area law behavior.  In ref. \cite{Kinar:1998vq} it was shown
 that this is achieved provided that either $g_{tt}g_{xx}(u)$ has a  non-vanishing minimum value or that it does not vanish at the value of
 the radial coordinate $u$ where $g_{tt}g_{uu}(u)\rightarrow \infty$, and where $g_{tt}, g_{xx}$ and $g_{uu}$ are the  metric components
 along   time, the  space direction of the Wilson line, and the radial direction respectively. Not surprisingly the  $AdS_5\times S^5$
 background  does not obey this requirement.

 A close cousin of this  background that does admit confinement is the  ``hard  wall model"
 discussed in the previous subsection  where one, by hand, chops off the radial direction to be $u\geq u_{\Lambda}$ where $u_\Lambda$
 is a scale in the bulk that corresponds to $\Lambda_{QCD}$ of the boundary confining gauge theory. This however is not a solution of the
 equations of motion.

 A prototype  confining background that is a solution  is the $AdS_5$ background with one spatial coordinate compactified \cite{Witten:1998zw}
 on a circle in such a way that the two dimensional manifold spanned by the radial direction and the circle has a cigar-like geometry. It is
 easy to check that upon imposing  anti-periodic boundary conditions for  fermions, the only massless fields of the dual large $N_c$ gauge
 theory are only the gauge fields and all their supersymmetric partners become massive. In that way supersymmetry is broken and the dual
 field theory is that of pure large $N_c$ gauge theory in three space-time dimensions. To get a gravity model dual of four dimensional
 confining large $N_c$ gauge theory, one can compactify the near horizon background of large number of $D4$ branes \cite{Itzhaki:1998dd}
 rather than the  $AdS_5\times S^5$ model  which is the background of   large number of $D3$ branes.
 In fact the dual gauge theory  is an  effective confining  theory with energies smaller than $\frac{1}{R}$ where $R$ is the radius of the compact
 circle   which maps into   the mass of the dual glueballs. There are several other solutions of the ten dimensional supergravity equations of
 motion that admit confinement but with no loss of generality we will discuss here only this model.

To incorporate in the gravity side the quark degrees of freedom one introduces $N_f$ flavor D-branes. For $N_f<<N_c$ one can neglect the
back-reaction of the flavor brane on the bulk and hence treat them as probe branes. In the Sakai Sugimoto model \cite{Sakai:2004cn} a stack
on $N_f$ D8 and a stack of $N_f$ anti- D8 branes are placed  so that asymptotically at large radial direction the transverse direction to their
worldvolumes is along the compact circle $x_4$.
 \begin{figure}[h!]
\begin{center}
\vspace{3ex}
\includegraphics[width= 100mm]{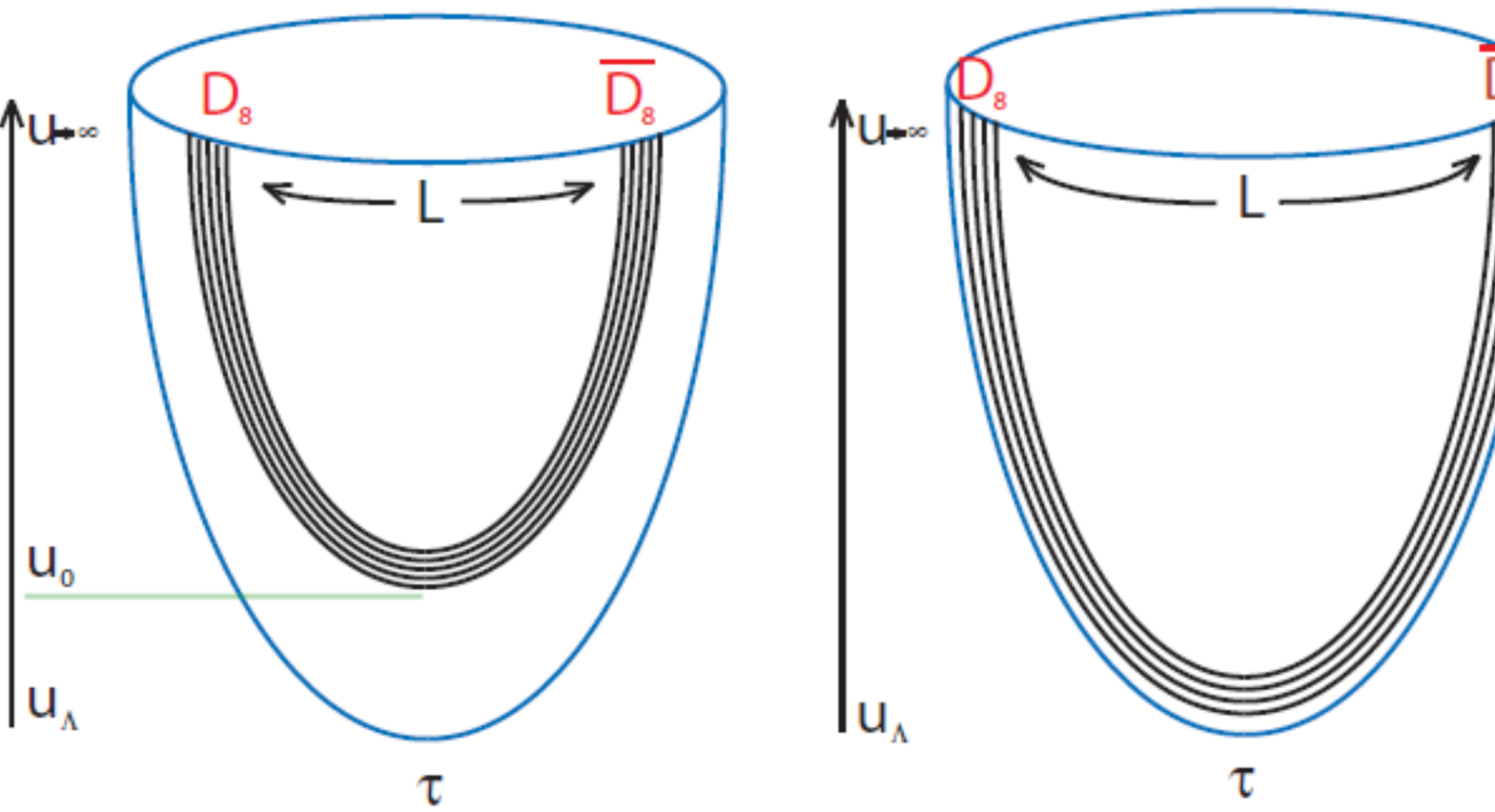}
\end{center}
\caption{ On the right side we have the antipodal set-up of the Sakai-Sugimoto model where $u_0=u_\Lambda$. On the left side we have
the generalized  non-antipodal set-up. }
\label{ssushape}
\end{figure}

 In the IR in the region of the tip of the cigar the two stacks of branes have to merge one into the other hence breaking the original
$U_L(N_f)\times U_R(N_f)$ chiral symmetry into a diagonal subgroup of $U_D(N_f)$. In original model, the U-shaped branes were
in an antipodal setup $u_0=u_\Lambda$, see the right  figure of (\ref{ssushape}).This was generalized  (see the left figure) to incorporate an
additional parameter $u_0\neq u_\Lambda$ \cite{Aharony:2006da}  which, as will be shown below,is crucial   for coupling the
protons to pions in the holographic picture.
The physics of the degrees of freedom that resides on the flavor branes, namely the $U(N_f)$ gauge fields and the scalars in
the adjoint of  the $U(N_f)$ group, is described by a DBI action. In fact the action includes, on top of the DBI action, also a CS
term. That is obviously where Heisenberg's model and holography meet. The action on the flavor branes in the Sakai-Sugimoto
model reads
\be\label{DBISS}
S_{DBI}= T_8\int d^{9}\sigma e^{-\tilde\phi} \sqrt{ - \det[\pa_\mu X^i\pa_\nu X^j g_{ij}(X) + 2\pi\alpha' F_{\mu\nu}]}\;,
\ee
where the dilaton $\phi$,  the metric  $g_{ij}$ and the RR four form are given by \cite{Aharony:2006da}
\begin{eqnarray}\label{S_S_metric}
ds^2\!\!\!\!&=&\!\!\!\bigg( \frac{u}{R_{D4}}\bigg)^{3/2}\!\bigg[\! \!-\!\!dt^2\!+\!\delta_{ij}dx^idx^j+f(u)dx_4^2\bigg]
\!+\!\bigg( \frac{R_{D4}}{u}\bigg)^{3/2}\!\bigg[\frac{du^2}{f(u)}\!+\!u^2d\Omega_4^2\bigg]\\ \nonumber
F_4\!&=&\!\frac{2\pi N_c}{V_4}\epsilon_4\ \ ,\ \ e^{\phi}=g_s\bigg( \frac{u}{R_{D4}}\bigg)^{3/4}
 , \ R_{D4}^3=\pi g_sN_cl_s^3 \ , \  \ f(u)=1-\bigg( \frac{u_{\Lambda}}{u}\bigg)^3\;,
\end{eqnarray}
where $x_4$ is the coordinate of the compactified circle, $V_4$ is the volume of the unit four sphere $\Omega_4$ and $\epsilon_4$ its
corresponding volume form.
Upon inserting the metric and the dilaton one finds, according to the general analysis in section 4.2,
\be\label{DBISSF}
S_{DBI}= \tilde T_8\int dt d^3 x dx_4 \; \phi^4 \sqrt{ f(\phi) + \bigg( \frac{R_{D4}}{\phi}\bigg)^{3} \left [\pa_\mu \phi \pa^\mu \phi
+ \frac{1}{f(\phi)} (\pa_{x_4} \phi)^2 \right ]}\;,
\ee
where $\tilde T_8=T_8\Omega_4/g_s$ and to connect to the rest of the paper we denoted the radial coordinate $u$ by $\phi$.


The fluctuations of $\phi$ translate using the dictionary of holography to scalar mesons.
To extract the spectrum of the latter one considers first a profile of the flavor brane given by $\phi_{cl}(x_4)$. One then introduces  the
fluctuations  of $\phi$ in the following form
\be
\phi(x_4,x^\mu)= \phi_{cl}(x_4) + \sum_n\delta\phi_n(x^\mu)\zeta_n(x_4).
\ee
The lowest mode  of the fluctuating field $\phi_0$ should be identified with the scalar field $\phi(x^\mu)$ in Heisenberg model.
Next one expands the DBI action to quadratic order in $\phi$, integrates over the $x_4$ direction, and derives a massive spectrum for the
$\delta\phi_n(x^\mu)$.
Here we do not want to expand the square root but rather maintain the full tower of derivatives of the field. The outcome of the integration of
the $\zeta_n(x_4)$ will be mass terms of the form $m_n^2\phi^2$ plus terms higher order in $\phi$. We assume here that the truncation to only
the mass term in the expansion of the DBI action can be translated to having a mass term in the  four dimenional DBI itself.
In that case the action takes the form
\be\label{DBISSM}
S_{DBI}= \tilde T_8\int dt d^3 x   \phi^4 \sqrt{ f(\phi) + \bigg( \frac{R_{D4}}{\phi}\bigg)^{3} \left [\pa_\mu \phi \pa^\mu \phi +  m^2\phi^2   \right ]}.
\ee
The equation of motion that associates with the action (\ref{DBISSM}) for the $m=0$ case can be written in the following form
\bea
&&\left(1-\bigg( \frac{u_{\Lambda}}{\phi}\bigg)^3\right)\left[ 8 - 5\bigg( \frac{u_{\Lambda}}{\phi}\bigg)^3 +\bigg( \frac{R_{D4}}{\phi}\bigg)^{3}
(\pa_\mu\phi\pa^\mu\phi - 2\phi\pa_\mu\pa^\mu \phi)\right ] \cr
&&+8\bigg( \frac{R_{D4}}{\phi}\bigg)^3\pa_\mu\phi\pa^\mu \phi -2\frac{u_\Lambda^3R_{D4}^3}{\phi^6}(\d_\mu\phi)^2-2\frac{R_{D4}^6}{\phi^6}
[(\d_\mu\phi)^2]^2\cr
&&+2\left(\frac{R_{D4}}{\phi}\right)^6\phi \d^\mu\phi[-(\d_\mu\phi)\d^\nu\d_\nu\phi
+(\d^\nu\phi)\d_\mu\d_\nu\phi]=0.
\eea
We leave the investigation of the relation between the solution of the DBI action given here and the DBI in
flat spacetime used in Heisenberg model to a future investigation.


\subsection{ A holographic description of the proton-proton scattering}

So far we have discussed the holographic laboratory and its relation to the DBI action applied in Heisenberg's model. Next we would like to
see what is the relation between  the cross section of a proton-proton scattering in Heisenberg's model and the corresponding cross section
in a holographic setup that associate with a confining theory equipped with flavor degress of freedom. Here for concreteness we will use the
generalized  Sakai-Sugimoto model. A  stringy realization of a baryon in this model \cite{Witten:1998xy}  is that  of a baryonic vertex made out
of a D4 brane that wraps the four cycle and is connected by $N_c$ strings to the $N_f$  probe flavor branes \cite{Brandhuber:1998xy}. A priori
the baryonic vertex could have been located in the generalized Sakai Sugimoto model in any place below the flavor brane, but in
\cite{Seki:2008mu} it was shown that in fact it must be immersed on the flavor brane. The interaction between two protons in this setup is
that of two baryonic vertices each connected to $N_c$ strings that stretch on the flavor branes. The scattering of such two objects is obviously
very complicated. Instead it was shown in \cite{Hata:2007mb} that one can view the baryon as a flavored gauge instanton. This follows from
the fact that the wrapped D4 brane is a point on the four dimensional part of the world-volume of the flavor brane which is spanned by the
ordinary three space coordinates and the radial direction.  Alternatively  it can be shown by expanding the flavor gauge DBI+ CS  actions,
keeping  only the leading order $U(N_f)$  YM + CS  action. The 5 dimensional action takes the form
\bea\label{YMCSaction}
S&=& S_{YM}+ S_{CS}\CR
S_{\rm YM}&\ \approx &\int\!\!d^4x\!\int\!\!dz\,\frac{1}{2g^2_{\rm YM}(z)}\,\tr\bigl({\cal F}_{MN}^2\bigr)\,,\CR
S_{\rm CS}\ &=&\ \frac{N_c}{16\pi^2}\int\! \hat A\wedge {\rm tr} F^2\
+\ \frac{N_c}{96\pi^2}\int\! A\wedge \hat{F}^2\;,
\eea
where near the bottom of the U-shaped flavor branes we have
\be
\frac{1}{2g^2_{\rm YM}(z)}\ =\ \frac{N_c\lambda M_{\rm KK}}{216\pi^3}\left( \zeta\,+\,\frac{8\zeta^3-5}{9\zeta}\,
M_{\rm KK}^2 z^2\, +\,O(M_{\rm KK}^4 z^4)\right).
\label{Uexpansion}
\ee
Here ${\cal F} $  is the $U(N_f)$ gauge field, and $A$ and  $\hat A$ denote  the gauge one-form associated with the $SU(N_f)$ and
$U(1)$ subgroups  respectively. We made a coordinate transformation from $(x^\mu,x_4)$ to  to a
 five-dimensional conformal  metric $
ds^2\ =\ \left(\frac{u(z)}{R_{D4}}\right)^{3/2}\bigl( -dt^2\,+\,dx_i^2\,+\,dz^2\bigr)\,,$
$\zeta=\frac{u_0}{u_\Lambda}$.
Based on  this action it was further shown   that the  static properties extracted from this model are similar to those derived from the Skyrme model.

Next we would like to examine to what extent does   Heisenberg's treatment of the  scattering of a proton on proton translate into a scattering
process of two instantons in the holographic laboratory. The interaction of the latter can be divided into three zones\cite{Hashimoto:2009ys}.
In the far zone when the distances between the two instantons is much larger than the inverse of the dual of $\Lambda_{QCD}$ the interaction
is dominated by the exchange of the lightest meson. In the isoscalar channel it was found out that the repulsion, due to the exchange of vector
mesons, is stronger than the attraction, due to the exchange of scalar mesons,  since the lightest meson on the latter type is heavier than the
lightest vector meson. In the isovector channel it is obvious that the lightest meson is the pion and the exchange of it yields an  attraction.
In the near zone, using the  solution that carries instanton number equal two, one finds that there is only a  repulsive
hard core interaction.  In the intermediate zone there is a repulsion due to the interaction of the instanton density with the $U(1)$ of the $U(N_f)$
flavor gauge group. However, as was shown in \cite{Kaplunovsky:2010eh} there is also an attractive force due to the interaction of the
instanton density with the scalar field associated with the fluctuation of the $D8$ branes.  The action of this scalar takes the form
\be\label{scalaractionPP}
S_\phi= S_{DBI} \quad+\ \frac{N_c}{16\pi^2}\int\!\!d^4x\!\int\!\!dz\ C(z)\times \tr\bigl(\Phi{\cal F}_{MN}{\cal F}^{MN}\bigr)\ +\ \cdots,
\ee
where  $C(z)$ measures the ratio of the attractive to the repulsive forces.
\be
\frac{F_a}{F_r}\
=\ C^2(z)\ =\ \frac{1-\zeta^{-3}}{9}\left(\frac{u_0}{u(z)}\right)^8\ \le\ \frac19\ <\ 1.
\label{ScalarToVector}
\ee
Note that for self-dual (instanton) configurations, $Tr[F_{MN} F^{MN}]=Tr[F\wedge F]$ and hence the scalar field that originates from the
brane fluctuations couples to the instanton density, namely to the  proton density.

  Thus, in a holographic description of the  interaction between   two protons,    both in the intermediate as well as in the far zone, the interaction
is mediated by a scalar field that is governed by a DBI action. The DBI action (\ref{DBISSF}) is not the one Heisenberg used  but rather a
DBI of a scalar in a curved background.  The source of the scalar field and its coupling to the proton  given in (\ref{scalaractionPP})
is different from the source of the scalar field discussed in section 5, but a fixed gauge field profile will generate a function $f(r)$ in the action as in
(\ref{sourceeq}), or an implicit external source as in (\ref{induction}).

\section{ Summary and open questions}

As was explained in the introduction, in this paper we addressed four aspects of Heisenberg's model of scattering of nucleons:

(i) We elaborated on, and gave further supporting evidence for the model. We made an analysis of the energy of the scalar field, and the
conditions under which we obtain the (almost) saturation of the Froissart bound. We have analyzed the what happens when we go from a
1+1 dimensional solution to a 3+1 dimensional one $\phi(s,r)$. We have understood the implicit source in the Heisenberg solution by analogy with the
electromagnetic Born-Infeld action: there is an "external" $\delta(s)$ source that is "spread out" by the medium. One can also consider $\delta(x^-)$
shockwave solutions by adding an explicit source in the Lagrangean. By using a perturbative $\phi(s,r)$, we have obtained corrections
away from the maximal Froissart saturation behaviour, as well as a new regime for $\sigma_{\rm tot}(s)$.

(ii) We examined the uniqueness of the DBI action in terms of giving the (almost) saturation of the bound. We have found that, perhaps surprisingly, no
action with a potential interaction, or with a finite number of higher derivative terms can do the job. The DBI action can do the job, though we have not been
able to prove that another action with infinite number of higher derivative terms cannot do as well.

(iii) Generalizations of the model. We proposed and analyzed several generalization of the Heisenberg model. We added a general potential inside
the square root, instead of just the mass term and we considered a sigma model with several scalar mesons.
We considered a "curved space" generalization inspired by holography, in particular the ``highly effective action" of \cite{Schwarz:2013wra}
for the case of single scalar in $AdS_5$, when we replace $\pa_\mu\phi\pa^\mu\phi$ by $\frac{1}{\phi^4} \pa_\mu\phi\pa^\mu\phi$, and
when considering the $n^{th}$ power rather than $\phi^4$ we have shown that only for the range $n\in (0,2)$ can we obtain saturation of the
bound. By considering a "black disk" type of amplitude in the sense of \cite{Kang:2004jd}, we have obtained also a value for the ratio
of the elastic to the total cross section, $\sigma_{\rm el}/\sigma_{\rm tot}$ that asymptotically goes to $1/4$. We note that the more common model
in for instance \cite{Block:1984ru} would give 1/2, but the experimental evidence points towards 1/4.

(iv) We have considered the relation of the Heisenberg model and the DBI action he considered to two holographic approaches to
proton-proton (or nucleon-nucleon) scattering: a simple hard-wall model, and a  more precise model based on flavor branes in confining backgrounds.

In this paper we have just explored the tip of the iceberg. There are  a handful of additional open questions that are awaiting
further investigation. Here we list few of them.
\begin{itemize}
\item
Probably the most interesting topic  related to realistic high energy scattering is performing a precise comparison between the results of
Heisenberg's model and experimental data of high energy scattering of nucleons and of nuclei. One can deduce the scattering total
cross section and the ratio between the elastic and total cross sections not only for the asymptotic range of energies as was discussed
in subsection 7.1. In section 4 we analyzed several possible generalizations of the model, and in section 8 we discussed the relation to
certain holographic models. These deviations from the original model can also be confronted with experimental data. One would like to
extract the values of the parameters of the various models that admit the best fit to the data. In particular the mass of the scalar
particle that mediates the interaction which we referred to as the ``pion" in this paper.
\item
It is well known that there are two approaches of phenomenological fitting the experimental data. One is based on the Froissart bound,
namely $\sigma_{tot}\sim \log^2(s)$ and the other on an exchange of Reggeons and Pomerons between the two scattering nucleons.
In this case one uses relation of the form  $\sigma_{tot}\sim a s^{-0.47} + b s^{0.08}$. Both approaches yield a reasonable fit (see
\cite{Nastase:2005bk} for a possible way to connect the gravity dual picture of gravitational shockwave scattering to the soft Pomeron behaviour).  Thus, a
natural question to ask is what is the relation between the two models. In section 8 we have attempted to relate the model to a holographic
model of scattering of nucleons. The latter is an approximated picture of a fully stringy description of the scattering process. The exchange
of a Reggeon and a Pomeron seem closely related to an exchange of an open and a closed string. Hence one may be able to find a direct
relation between the two approaches.
\item
One natural generalization of the model that was not discussed here but in fact is quite common in implementing the DBI action in
holography is the non-abelian DBI model. To incorporate the (flavor) non-abelian nature of the pions is the analog of using $N_f$
probe flavor branes rather than a single one in holographic models. In both cases the non-abelianization will provide further structure. A
first try for the nonabelian model was presented in \cite{Nastase:2005pb}.
\item
Describing the scattering of two nucleons as a scattering of two shockwaves is clearly only an approximation and one may attempt at
introducing correction beyond the shockwave limit. Similarly one can introduce corrections to the black disk model.

\end{itemize}






{\bf Acknowledgements}. JS would like   to thank S. Nussinov and O. Oz for useful discussions.
The research of HN is supported in part by CNPQ grant 301709/2013-0
and FAPESP grant 2013/14152-7. The work of JS was supported in part by
a centre of excellence supported by the Israel
Science Foundation (grant number 1989/14), and  by
 the US-Israel bi-national fund (BSF) grant number 2012383 and the Germany–Israel bi-national fund GIF grant number I-244-303.7-2013.

\appendix

\section{ An alternative method of determining the scalar field energy}

The Hamiltonian density was given in (\ref{Hamden}). It reads
 \be
{\cal H}=p\dot\phi-{\cal
L}=\frac{l^{-4}+(\nabla\phi)^2+m^2\phi^2}{\sqrt{1+l^4[(\d_\mu\phi)^2+m^2\phi^2]}}-l^{-4}\;,\label{HamdenA}
\ee

 To determine the Hamiltonian density in momentum space namely ${\cal H}(k)$ is a non-trivial task for the DBI action since  we
 cannot simply, as is done for ordinary free field theories, substitute the Fourier transform of the field into (\ref{HamdenA}) since the fields
 appear also in the denominator.  In  case that upon substituting the classical solution $\phi(s)$ into the denominator the latter is a constant
 then one can use the usual method. But  as was shown in section 2  this is not the case for the DBI action and hence  one has to adopt
 another approach. Here we suggest such an alternative.
 Define now the Fourier transform of ${\cal H}$ as follows
 \be
\sqrt{{\cal H}(x,t)} = \int \frac{dk}{\sqrt{2\pi}} \frac{[e^{-ikx} \tilde{\cal F}(k,t) + e^{ikx} \tilde{\cal F}^*(k,t)]}{2}
 \ee
 and substitute it into the energy, so that
 \bea\label{EFk}
E &=& \frac{1}{2\pi}\int dx  \int dk \frac{[e^{-ikx} \tilde{\cal F}(k,t) + e^{ikx} \tilde{\cal F}^*(k,t)]}{2}\int dp
\frac{[e^{-ipx} \tilde{\cal F}(p,t) + e^{ipx} \tilde{\cal F}^*(p,t)]}{2}\cr
&=& \int dk \frac{[2\tilde{\cal F}(k,t)\tilde{\cal F}^*(k,t)+\tilde{\cal F}(k,t)\tilde{\cal F}(-k,t)+\tilde{\cal F}^*(k,t)\tilde{\cal F}^*(-k,t)]}{4}
 \eea
For the theory of a  free massless scalar  in two space-time dimensions ${{\cal H}(x,t)}= \frac12[(\pa_x\phi)^2 + (\pa_0\phi)^2]$.
In this case it is easy to see that $\tilde{\cal F}(k,t)= \sqrt{k} a(k)$ where the field $\phi(x,t)$ has a Fourier transform
$\phi(x,t) = \int dk \frac{1}{\sqrt{2\pi k}}[ a(k) e^{-ikx}+ a^\dagger(k) e^{+ikx}]$. In the case of massive free scalar field we get
$\tilde{\cal F}(k,t)= \sqrt{\sqrt{k^2+m^2} } a(k)$. In these cases the only contributions to (\ref{EFk})will be from the
$\tilde{\cal F}(k,t)\tilde{\cal F}^*(k,t)$ term.  For the general case one has to first determine $\tilde{\cal F}(k,t)$ and then $E$ is given by (\ref{EFk}).

Following this approach we now have to find the Fourier transform of $\sqrt{{\cal H}(s)}$. We cannot find an exact analytic expression for
 it neither for the massless case nor for the massive one. From the analysis of the energy as an integral over $s$ one finds 
 that the main contribution to the energy comes from the region of small $s$. Thus we can get an approximation of the dependence of
 the energy on $\gamma$ using the  the leading order in $s$ expression of
 $\sqrt{{\cal H}(s)}\sim \frac{1}{l^2 \sqrt{m}}\frac{t}{s^{3/4}}=\frac{1}{l^2 \sqrt{m}}\frac{t}{(t^2-x^2)^{3/4}} $. Its Fourier transform reads
\be
TF\left[ \sqrt{{\cal H}(s)}\right ]\sim \frac{1}{l^2 \sqrt{m}}\frac{\sqrt[4]{2} \sqrt[4]{\left| k\right| } K_{-\frac{1}{4}}\left(\frac{\left| k\right|
   }{\sqrt{-\frac{1}{t^2}}}\right)}{\left(-\frac{1}{t^2}\right)^{5/8}
   \left(t^2\right)^{3/4} \Gamma \left(\frac{3}{4}\right)}.
\ee
Expanding this expression in $\frac{1}{k}$ we get
\be
\frac{\sqrt{\pi } e^{-\frac{\left| k\right| }{\sqrt{-\frac{1}{t^2}}}} \left(1-\frac{3
   \sqrt{-\frac{1}{t^2}}}{32 \left| k\right| }\right)}{\sqrt[4]{2}
   \left(-\frac{1}{t^2}\right)^{3/8} \left(t^2\right)^{3/4} \Gamma
   \left(\frac{3}{4}\right) \sqrt[4]{\left| k\right| }}.
\ee
Substituting this expression in the energy and taking the integration region to be $\gamma m> k> m$ we finally get that
\be
E\sim \sqrt{\gamma m}.
\ee
The reason that this result  does not match the result found in section 2 is that we took a crude approximation of $\sqrt{{\cal H}(s)}$.
Obviously this approximation can be systematically improved by improving the approximation of $\sqrt{{\cal H}(s)}$.

\section{Scalar solutions in $0+1$ dimensions}

Here for completeness we write down solutions of the Heisenberg action in 0+1 dimensions. The equations of motion in this case are
\be
\ddot{\phi}+ m^2 \phi + l (\dot\phi)^2\frac{\ddot{\phi}-m^2 \phi}{1-l[(\dot\phi)^2- m^2\phi^2]}=0.
\ee
For the massless case the equation reduces to $\ddot{\phi}=0$ and hence the solution takes the form $\phi= a t +b$.
For the massive case  the solution takes the form
\bea
y(x)&=& \frac{i \text{sn}\left(i m \sqrt{l c_1+1} x+i m \sqrt{l
   c_1+1} c_2|\frac{l c_1}{l c_1+1}\right)}{\sqrt{l}
   m},\cr
y(x)&=&  -\frac{i \text{sn}\left(i \left(m \sqrt{l c_1+1}
   x+m \sqrt{l c_1+1} c_2\right)|\frac{l c_1}{l c_1+1}\right)}{\sqrt{l}
m}.
\eea

The solution  takes the following form for various values of $\frac{m}{l}=0.1,1,5$

\begin{figure}[h!]
\begin{center}
\vspace{3ex}
\includegraphics[width= 100mm]{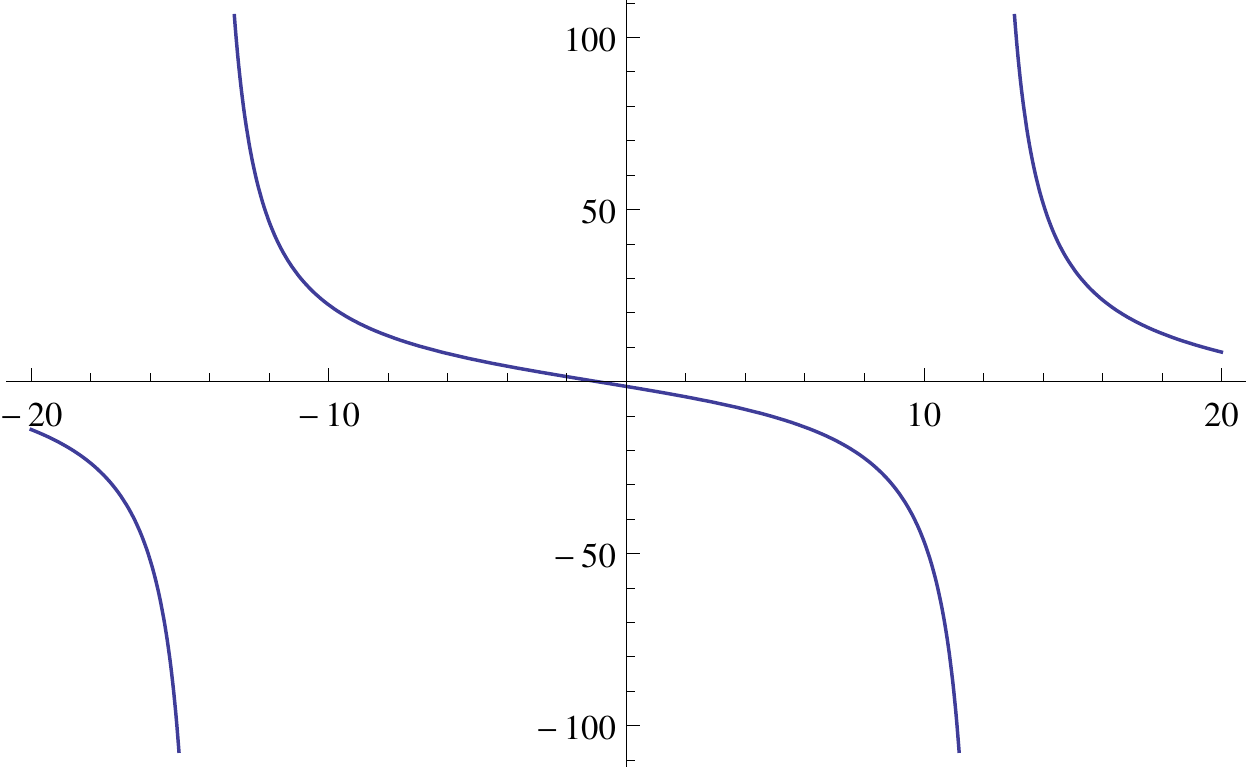}

\end{center}
\caption{$\phi(t)$ as a function of $t$ for$\frac{m}{l}=0.1$ }
\label{Self}
\end{figure}
\begin{figure}[h!]
\begin{center}
\vspace{3ex}
\includegraphics[width= 100mm]{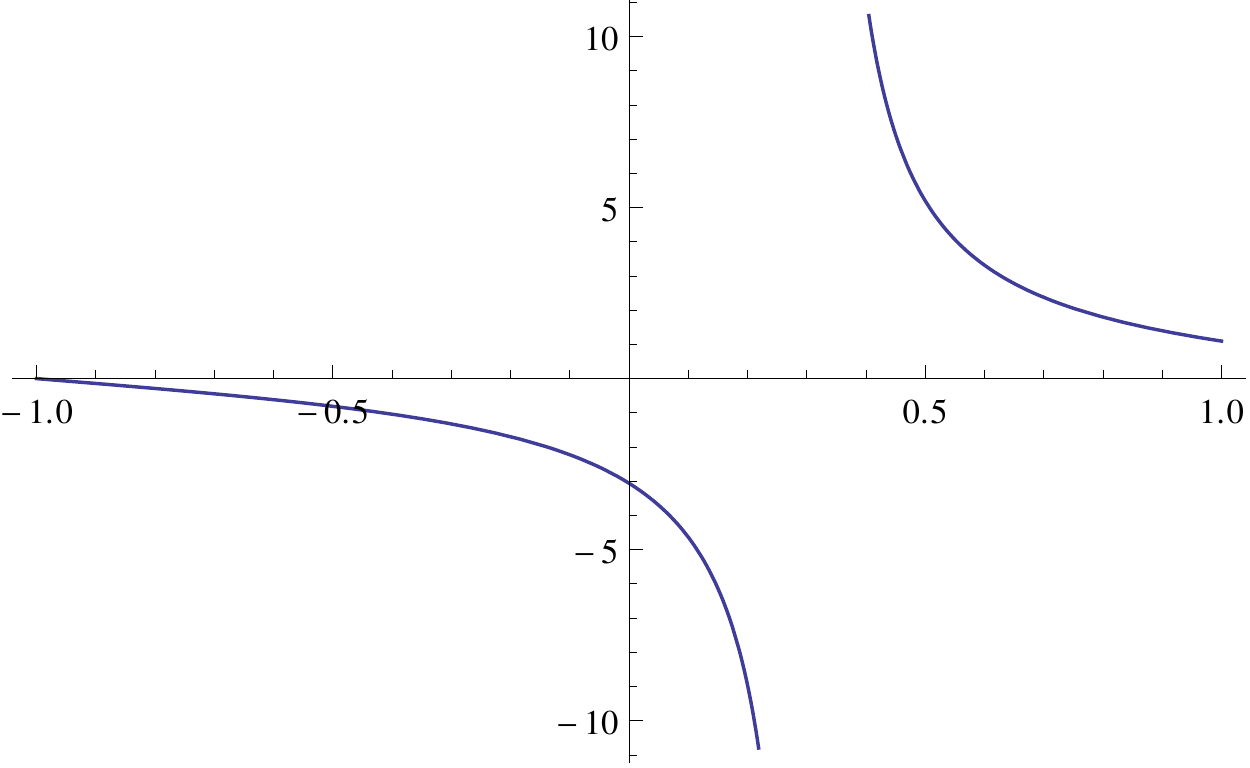}
\end{center}
\caption{{$\phi(t)$ as a function of $t$ for$\frac{m}{l}=1$ }}
\label{Self2}
\end{figure}
\begin{figure}[h!]
\begin{center}
\vspace{3ex}
\includegraphics[width= 100mm]{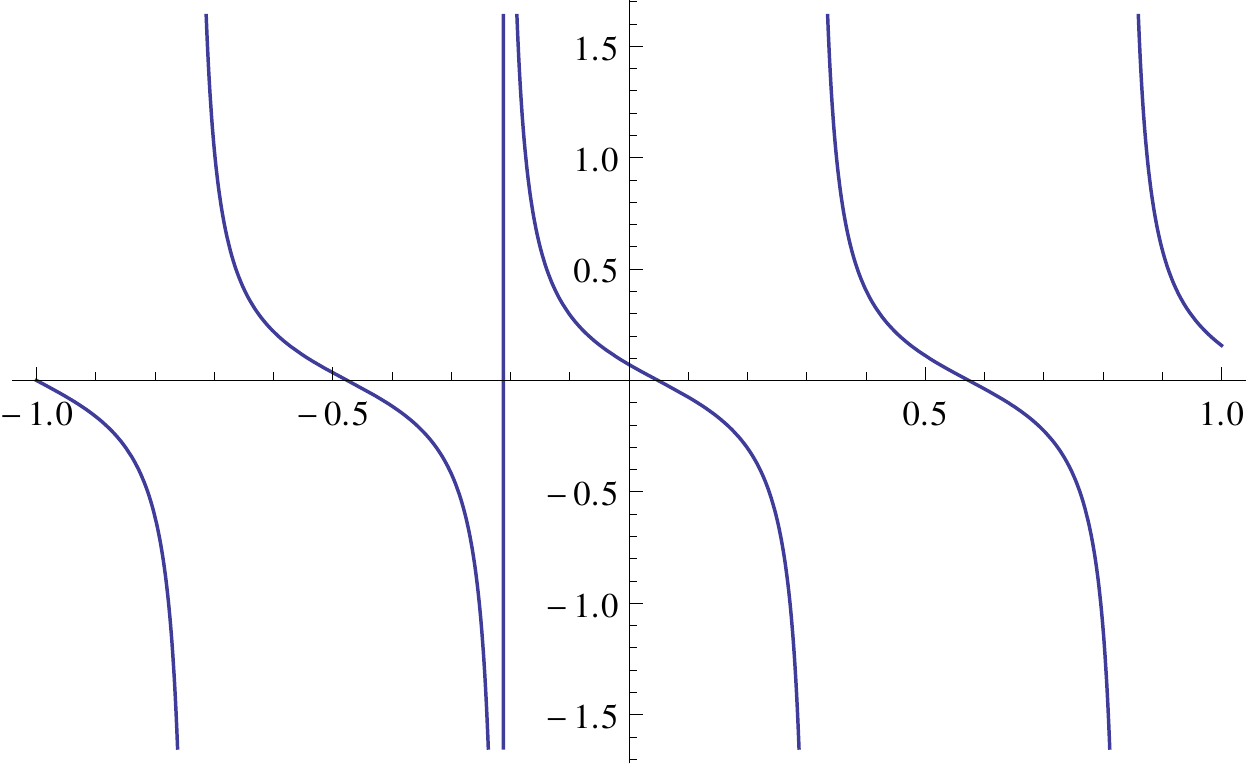}
\end{center}
\caption{{$\phi(t)$ as a function of $t$ for$\frac{m}{l}=5$ }}
\label{Self3}
\end{figure}

For the one dimensional case the Hamiltonian density (\ref{Hamden}) is the Hamiltonian and hence we can write a first order differential
equation  which is its conservation in time instead of the equation of motion.  The Hamiltonian for this case  reads
\be
  H l = \frac{1}{\sqrt{1-l [(\dot\phi)^2-m^2\phi^2]}}[1+lm^2\phi^2)] -1.
\ee
Thus  the  first order differential equation is
\be
(\dot\phi)^2 = \frac{H(lH+2)}{(lH+1)^2} + m^2 \left( 1-\frac{2}{(lH+1)^2}\right)\phi^2 -\frac{l m^4\phi^4}{(lH+1)^2}\;,
\ee
or in an integral form
\be
\int  \frac{d\phi}{\sqrt{ \frac{H(lH+2)}{(lH+1)^2} + m^2 ( 1-\frac{2}{(lH+1)^2})\phi^2 -\frac{l m^4\phi^4}{(lH+1)^2}}} = t+ c.
\ee

\section{Scalar solutions in 1+1 dimensions: static and depending independently on $x^+$ and $x^-$.}

Before discussing a genuine two dimensional case let's check the equation for a (soliton) static solution.
For that case the equation takes the form
\be
\pa^2_x{\phi}- m^2 \phi - l^2 (\pa_x\phi)^2\frac{\pa_x^2{\phi}+m^2 \phi}{1+l^2[(\pa_x\phi)^2+ m^2\phi^2]}=0.
\ee
This equation admits an analytic solution similar to the one of the one dimensional case, namely
\bea
y(x)&=& \frac{i \text{sn}\left(i m \sqrt{l^2 c_1-1} x+i m
   \sqrt{l^2 c_1-1} c_2|\frac{l^2 c_1}{l^2 c_1-1}\right)}{l
   m} \cr
y(x)&=&  -\frac{i \text{sn}\left(i \left(m \sqrt{l^2
c_1-1} x+m \sqrt{l^2 c_1-1} c_2\right)|\frac{l^2 c_1}{l^2 c_1-1}\right)}{lm}.
\eea
This soliton solution is similar to the solution of the one dimensional case discussed above.
It is to see from the equations of motion that the map $l\rightarrow -l^2$ and $m^2\rightarrow -m^2$
maps the one dimensional equation to the solitonic two dimensional one.

We next consider the truly two dimensional case, thought of as an approximation for the
four dimensional system of colliding shock waves in the limit of zero width for the shock wave,
and in a limit of azimuthal symmetry in plane of the shock. It is convenient in two dimensions to use light-cone coordinates
$x^\pm=t\pm x$, with
\be
\pa_+=\pa_{x^+}=\frac12 (\pa_t+\pa_x) \qquad \pa_-=\pa_{x^-}=\frac12 (\pa_t-\pa_x).
\ee
In these light cone coordinates
\be
\pa_\mu\phi\pa^\mu\phi = (\dot\phi)^2-(\phi')^2 = 4\pa_+\phi\pa_-\phi  \qquad  \pa_\mu\pa^\mu\phi =\ddot \phi-\phi''  = 4\pa_+\pa_-\phi.
\ee
We now define the following coordinates
\be
s = t^2-x^2 = x^+x^-  \qquad q=\frac{x^-}{x^+}.
\ee
For these coordinates we find that
\be\label{dfdf}
\pa_\mu\phi\pa^\mu\phi = +4\left[s(\pa_s\phi)^2 - \frac{q^2}{s}(\pa_q\phi)^2\right] \qquad  \pa_\mu\pa^\mu\phi =
+4[\pa_s(s\pa_s\phi) -\frac{q}{s}[\pa_q(q\pa_q\phi)]\;,
\ee
and also
\bea\label{ddfdf}
&&(2(\d_\mu\phi)(\d_\nu\phi)(\d^\mu\d^\nu\phi)=)\pa_\mu\phi\pa^\mu(\pa_\nu\phi\pa^\nu\phi)= 16[s(\pa_s\phi)^3 + 2s^2(\pa_s\phi)^2\pa_s^2\phi +\cr
&&\frac{q^2}{s}(\pa_q\phi)^2(\pa_s\phi)-2q^2 (\pa_s\phi)(\pa_q\phi)(\pa_q\pa_s\phi) +2\frac{q^3}{s^2}(\pa_q\phi)^3
+2\frac{q^4}{s^2}(\pa_q\phi)^2(\pa^2_q\phi) ].
\eea
Substituting (\ref{dfdf}) and (\ref{ddfdf}) into the equation of motion (\ref{dbieom}) we get that  for the variables $s$ and $q$ the
equation of motion takes the form
\bea
&& 4[\pa_s(s\pa_s\phi) -\frac{q}{s}\pa_q\phi - \frac{q^2}{s}\pa_q^2\phi] +m^2 \phi = \cr
&&4 l^4m^2\phi \frac{[s(\pa_s\phi)^2 - \frac{q^2}{s}(\pa_q\phi)^2]}{1+l^4[m^2\phi^2 -4[s(\pa_s\phi)^2 - \frac{q^2}{s}(\pa_q\phi)^2]]} \cr
&&-8l^4  \frac{[s(\pa_s\phi)^3 + 2s^2(\pa_s\phi)^2\pa_s^2\phi+
\frac{q^2}{s}(\pa_q\phi)^2(\pa_s\phi)}{1+l^4[m^2\phi^2 -4[s(\pa_s\phi)^2 - \frac{q^2}{s}(\pa_q\phi)^2]]}\cr
&&
+\frac{-2q^2 (\pa_s\phi)(\pa_q\phi)(\pa_q\pa_s\phi) +2\frac{q^3}{s^2}(\pa_q\phi)^3 +2\frac{q^4}{s^2}(\pa_q\phi)^2(\pa^2_q\phi)}{1+l^4[m^2\phi^2
-4[s(\pa_s\phi)^2 - \frac{q^2}{s}(\pa_q\phi)^2]]}. \cr
&&
\eea
For the special case of $\phi(s,q)=\phi(s)$ the equation of motion reduces to
\be
4\pa_s(s\pa_s\phi) +m^2 \phi =
 4 l^4s(\pa_s\phi)^2\frac{m^2\phi +2[(\pa_s\phi) + 2s\pa^2_s\phi ]}{1+l^4[m^2\phi^2 -4[s(\pa_s\phi)^2 ]]}\;,
\ee
which is the same as (\ref{question}), so it reduces to the equation of motion of Heisenberg.

\newpage

\bibliographystyle{utphys}
\bibliography{heisenberg}

\end{document}